\documentclass[aps,prd,superscriptaddress,showpacs,preprintnumbers,floatfix]{revtex4}
\usepackage{graphicx,color}

\newcommand{\be}{\begin{equation}}
\newcommand{\ee}{\end{equation}}
\newcommand{\bea}{\begin{eqnarray}}
\newcommand{\eea}{\end{eqnarray}}

\newcommand{\bfk}{\mbox{\boldmath $k$}}

\newcommand{\bfp}{\mbox{\boldmath $p$}}

\newcommand{\bfP}{\mbox{\boldmath $P$}}

\newcommand{\bfx}{\mbox{\boldmath $x$}}

\newcommand{\bfDelta}{\mbox{\boldmath $\Delta$}}
\newcommand{\bfpartial}{\mbox{\boldmath $\partial$}}

\newcommand{\avkt}{\langle k_\perp^2 \rangle}
\def\lsim{\mathrel{\rlap{\lower4pt\hbox{\hskip1pt$\sim$}}\raise1pt\hbox{$<$}}}
\def\gsim{\mathrel{\rlap{\lower4pt\hbox{\hskip1pt$\sim$}}\raise1pt\hbox{$>$}}}
\def\nostrocostruttino#1\over#2{\mathrel{\mathop{\kern 0pt \rlap
{\hbox{$#1$}}} \hbox{\kern-.135em $#2$}}}

\def\kt{k_\perp}

\def\bkt{\bfk_\perp}

\def\pp{p_\perp}

\def\xb{x_{_{\!B}}}

\def\avk{\langle k_\perp ^2\rangle}
\def\avp{\langle p_\perp ^2\rangle}
\def\avPT{\langle P_T^2\rangle}

\begin{document}
%
\preprint{JLAB-THY-13-1830} 
\title{Unpolarised Transverse Momentum Dependent Distribution and Fragmentation Functions from SIDIS Multiplicities}
\author{M.~Anselmino}
\affiliation{Dipartimento di Fisica Teorica, Universit\`a di Torino,
             Via P.~Giuria 1, I-10125 Torino, Italy}
\affiliation{INFN, Sezione di Torino, Via P.~Giuria 1, I-10125 Torino, Italy}
\author{M.~Boglione}
\affiliation{Dipartimento di Fisica Teorica, Universit\`a di Torino,
             Via P.~Giuria 1, I-10125 Torino, Italy}
\affiliation{INFN, Sezione di Torino, Via P.~Giuria 1, I-10125 Torino, Italy}
\author{J.O.~Gonzalez~H.}
\affiliation{INFN, Sezione di Torino, Via P.~Giuria 1, I-10125 Torino, Italy}
\author{S.~Melis}
\affiliation{Dipartimento di Fisica Teorica, Universit\`a di Torino,
             Via P.~Giuria 1, I-10125 Torino, Italy}
\affiliation{INFN, Sezione di Torino, Via P.~Giuria 1, I-10125 Torino, Italy}
\author{A.~Prokudin}
\affiliation{Jefferson Laboratory, 12000 Jefferson Avenue, Newport News, VA 23606, USA}
\date{\today}

\begin{abstract}
The unpolarised transverse momentum dependent distribution and fragmentation 
functions are extracted from HERMES and COMPASS experimental measurements 
of SIDIS multiplicities for charged hadron production. The data are grouped 
into independent bins of the kinematical variables, in which the TMD 
factorisation is expected to hold. A simple factorised functional form of the 
TMDs is adopted, with a Gaussian dependence on the intrinsic transverse 
momentum, which turns out to be quite adequate in shape. HERMES data do not 
need any normalisation correction, while fits of the COMPASS data much improve 
with a $y$-dependent overall normalisation factor. A comparison of the extracted 
TMDs with previous EMC and JLab data confirms the adequacy of the simple
gaussian distributions. The possible role of the TMD evolution is briefly 
considered. 
\noindent
\end{abstract}


\maketitle

\section{\label{Intro} Introduction}

Tranverse momentum dependent parton densities (TMD-PDFs) and fragmentation 
functions (TMD-FFs), often collectively referred to as TMDs, have recently 
attracted and keep receiving a huge amount of interest. TMD-PDFs show remarkable 
spin and angular momentum correlation properties of quarks and gluons and allow 
a direct connection to the internal 3-D partonic structure of 
hadrons~\cite{Boer:2011fh}. TMD-FFs reveal universal features of the hadronization
process and couple to TMD-PDFs in several physical observables.  

Studies of TMDs are mostly performed in polarised Semi Inclusive Deep Inelastic 
Scattering (SIDIS) processes, $\ell \, N \to \ell ^\prime \, h \, X$, in the framework of 
the TMD factorisation scheme~\cite{Ji:2004xq,Ji:2004wu,Bacchetta:2008xw,Collins:2011zzd}, 
according to which the SIDIS cross section is written as a convolution 
of TMD-PDFs, TMD-FFs and known elementary interactions. The role of TMDs in such 
processes has been definitely established by the observation and interpretation 
of single spin asymmetries, which confirmed the existence of polarised TMDs like 
the Sivers distribution~\cite{Sivers:1989cc,Sivers:1990fh} and the Collins 
fragmentation function~\cite{Collins:1992kk}. 

The role of TMDs is already evident in unpolarised cross sections, simply by 
looking at the transverse momentum, $P_T$, distribution of the final hadron in 
the $\gamma^*-N$ centre of mass frame, or, at order $1/Q$, at the azimuthal 
dependence of the hadron around the $\gamma^*$ direction. 
In Ref.~\cite{Anselmino:2005nn} a first investigation of SIDIS unpolarised cross 
sections was performed, mainly based on the EMC Collaboration experimental 
data~\cite{Ashman:1991cj}, gathered from SIDIS experiments at different energies 
and off different targets. This analysis assumed a simple factorised and Gaussian 
parameterisation of the TMDs, which was later confirmed by the independent study 
of Ref.~\cite{Schweitzer:2010tt}, based on data from 
JLab~\cite{Osipenko:2008aa,Mkrtchyan:2007sr} and HERMES~\cite{Airapetian:2009jy}.  

However, very recently, plenty of new data on SIDIS multiplicities 
have been made available by the HERMES~\cite{Airapetian:2012ki} and 
COMPASS~\cite{Adolph:2013stb} Collaborations. Both of them have performed 
multivariate analyses of their measurements, providing extremely rich data sets, 
arranged in independent bins of the kinematical variables, which give a unique 
opportunity to extract new and more detailed information on the unpolarised TMDs. 
Also the theoretical investigation of TMDs has much progressed 
lately~\cite{Ji:2004xq,Ji:2004wu,Collins:2011zzd,Aybat:2011zv,Aybat:2011ge,
Echevarria:2012pw,Echevarria:2012qe} with the study of the QCD evolution of the 
Sivers and unpolarised TMDs \mbox{-- the so-called TMD evolution --} with the first 
phenomenological applications~\cite{Aybat:2011ta,Anselmino:2012aa,
Bacchetta:2013pqa,Godbole:2013bca,Sun:2013dya,Sun:2013hua,Boer:2013zca}.

In this paper we analyse the latest HERMES and COMPASS data on unpolarised 
multiplicities aiming at improving our knowledge of the unpolarised TMDs. 
We reconsider, with the support of the new data, the first extraction of 
Ref.~\cite{Anselmino:2005nn}; somewhat surprisingly, it turns out that the 
simple factorised form of the TMDs with the original, flavour independent, 
Gaussian parameterisation, still works rather well. However, the observed 
(Gaussian) dependence of the SIDIS cross section on the hadron transverse
momentum, $P_T$, is generated by a combination of the (Gaussian) dependences 
in the quark TMD-PDF and TMD-FF; thus, it is rather difficult to fix separately 
the parameters of the two Gaussians by studying only unpolarised multiplicities.

Although the HERMES and COMPASS data cover similar $Q^2$ regions 
($1 \le Q^2 \le 10 $ GeV$^2$), they differ in the experimental set-up, 
in the statistics, in the binning choices and in the explored $\xb$ range;
in addition, there seems to be some discrepancy between the two measurements. 
We then fit the HERMES and the COMPASS multiplicities separately.
A simultaneous fit of both sets of data would lead to poor results and is not 
presented here.
 
Recently, another study of the unpolarised TMDs has 
appeared~\cite{Signori:2013mda}, which follows a procedure somehow similar to 
that of this work, but which considers only the HERMES set of experimental 
data and does not include any attempt to check for signs of scale evolution. 

After a short Section II devoted to the formalism, we present our main results 
in Section III. In Section IV we briefly discuss the possible role, and look 
for possible signs, of TMD evolution.
In Section V we compare our present results with those of previous 
analyses~\cite{Anselmino:2005nn, Schweitzer:2010tt} and check their 
consistency with other measurements of SIDIS cross sections and 
$P_T$-distributions~\cite{Ashman:1991cj, Osipenko:2008aa, Mkrtchyan:2007sr, 
Asaturyan:2011mq} which were not included in our fits. Further comments and 
concluding discussions are presented in Section VI.

\section{\label{Form} Formalism}

The unpolarised $\ell \, + \, p  \to \ell^\prime \, h \, X$, SIDIS cross section 
in the TMD factorisation scheme, at order $(k_\perp/Q)$  and $\alpha_s^0$,
in the kinematical region where $P_T \simeq k_\perp \ll Q \>$, 
reads~\cite{Bacchetta:2006tn, Anselmino:2011ch}:
\bea
\frac{d\sigma^{\ell + p \to \ell^\prime h X}}
{d\xb \, dQ^2 \, dz_h \, dP_T^2} &=&
\frac {2 \, \pi^2 \alpha^2}{(\xb s)^2} \, \frac{\left[ 1 + (1-y)^2 \right]}{y^2}
\nonumber \\
&\times& \sum_{q} e_q^2 \,
\int d^2\bkt \, d^2\bfp_\perp
\> \delta ^{(2)}\Big(\bfP_T - z_h\bfk_\perp -\bfp_\perp\Big)\,
f_{q/p} (x, k_{\perp}) \, D_{h/q}(z, p_{\perp}) \label{xs-full-expl} \\
&\equiv& \frac{2 \, \pi^2 \alpha^2}{(\xb s)^2} \, 
\frac{\left[1 + (1-y)^2 \right]}{y^2} \, F_{UU} \> \cdot \nonumber 
\eea
In the $\gamma^* - p$ c.m.~frame the measured transverse momentum,
$\bfP_T$, of the final hadron is generated by the transverse momentum of the
quark in the target proton, $\bfk_\perp$, and of the final hadron with respect
to the fragmenting quark, $\bfp_\perp$. At order $k_\perp/Q$ it is simply given by
\be
\bfP_T =  z \,\bfk_\perp + \bfp_\perp \>.
\ee
As usual:
\be
s = (\ell + p)^2 \quad\quad Q^2=-q^2 = -(\ell -\ell')^2 \quad\quad
\xb = \frac {Q^2}{2p \cdot q}\quad\quad y = \frac {Q^2}{\xb s} \quad\quad
z_h = \frac{p \cdot P_h}{p \cdot q} \> 
\ee
and the variables $x$, $z$ and $\bfp _\perp$ are related to the final observed 
variables $\xb$, $z_h$ and $\bfP_T$ and to the integration variable $\bfk_\perp$. 
The exact relations can be found in Ref.~\cite{Anselmino:2005nn};
at ${\cal O}(k_\perp/Q)$ one simply has
\be
x = \xb \quad\quad z = z_h 
\>.
\ee

The unpolarised TMD distribution and fragmentation functions, 
$f_{q/p} (x, k_{\perp})$ and $D_{h/q}(z, p_{\perp})$, depend on the light-cone 
momentum fractions $x$ and $z$ and on the magnitudes of the transverse momenta 
$\kt = |\bfk_\perp|$ and $\pp = |\bfp_\perp|$. We assume these dependences
to be factorized and we assume for the $\kt$ and $\pp$ dependences a Gaussian 
form, with one free parameter which fixes the Gaussian width,
\bea
&&f_{q/p} (x,\kt)= f_{q/p} (x)\,\frac{e^{-\kt^2/\avk}}{\pi\avk}
\label{unp-dist}\\
&&D_{h/q}(z,\pp)=D_{h/q}(z)\,\frac{e^{-\pp^2/\avp}}{\pi\avp}\,\cdot
\label{unp-frag}
\eea
The integrated PDFs, $f_{q/p}(x)$ and $D_{h/q}(z)$, can be taken from the 
available fits of the world data: in this analysis we will use the CTEQ6L set 
for the PDFs~\cite{Pumplin:2002vw} and the DSS set for the fragmentation 
functions~\cite{deFlorian:2007aj}. In general, the widths of the Gaussians could 
depend on $x$ or $z$ and might be different for different distributions: here, 
we first assume them to be constant and flavour independent and then perform
further tests to check their sensitivity to flavour, $x$, $z$ and $Q^2$ 
dependence. The constant Gaussian parameterisation, supported by a number of 
experimental evidences~\cite{Schweitzer:2010tt} as well as by dedicated lattice 
simulations~\cite{Musch:2007ya}, has the advantage that the intrinsic transverse 
momentum dependence of the cross section can be integrated out analytically. 
In fact, inserting Eqs.~(\ref{unp-dist}) and (\ref{unp-frag}) into 
Eq.~(\ref{xs-full-expl}), one obtains 
\be
F_{UU}  =  \sum_{q} \, e_q^2 \,f_{q/p}(\xb)\,D_{h/q}(z_h)
\frac{e^{-P_T^2/\avPT}}{\pi\avPT} \label{G-FUU}
\ee
where
\be
\avPT = \avp + z_h^2 \, \avk \>. \label{avPT}
\ee
Notice that $\avk$ and $\avp$ will be taken as the free parameters of our fit.

According to COMPASS~\cite{Adolph:2013stb} notation the differential 
hadron multiplicity is defined as:
\be
\frac{d^2 n^h(\xb, Q^2, z_h, P_T^2)}{dz_h \, dP_T^2} \equiv
\frac{1}{\displaystyle{\frac{d^2 \sigma^{DIS} (\xb, Q^2)}{d\xb \, dQ^2}}} \>
\frac{d^4 \sigma (\xb, Q^2, z_h, P_T^2)}{d\xb \, dQ^2 \, dz_h \, dP_T^2} \>,
\label{mult-c}
\ee
while HERMES~\cite{Airapetian:2012ki} definition is 
\be
M_n^h(\xb, Q^2,z_h, P_T)
\equiv
\frac{1}{\displaystyle{\frac{d^2 \sigma^{DIS} (\xb, Q^2)}{d\xb \, dQ^2}}} \>
\frac{d^4 \sigma (\xb, Q^2, z_h, P_T)}{d\xb \, dQ^2 \, dz_h \, dP_T}
\> \cdot \label{mult-h}
\ee
where the index $n$ denotes the kind of target. 

The Deep Inelastic Scattering (DIS) cross section has the usual leading 
order collinear expression,  
\be
\frac{d^2 \sigma^{DIS} (\xb, Q^2)}{d\xb \, dQ^2} = 
\frac {2 \, \pi \, \alpha^2}{(\xb s)^2} \, \frac{\left[ 1 + (1-y)^2 \right]}{y^2}
\sum_{q} e_q^2 \> f_{q/p} (\xb) \label{xs-DIS} \> \cdot
\ee

Inserting Eq.~(\ref{xs-full-expl}), (\ref{G-FUU}) and (\ref{xs-DIS}) into 
Eq.~(\ref{mult-c}) we have a simple explicit expression for the COMPASS and 
HERMES multiplicities: 
\be
\frac{d^2 n^h(\xb, Q^2, z_h, P_T)}{dz_h \, dP_T^2} = 
\frac{1}{2P_T} M_n^h(\xb, Q^2,z_h, P_T) =
\frac{\pi \> \sum_{q} \, e_q^2 \,f_{q/p}(\xb)\,D_{h/q}(z_h)}
{\sum_{q} e_q^2 \> f_{q/p} (\xb)} \> 
\frac{e^{-P_T^2/\avPT}}{\pi\avPT}
\>, \label{mult-gaus}
\ee
with $\avPT$ given in Eq.~(\ref{avPT}). Notice that, by integrating the above 
equation over $\bfP_T$, with its magnitude ranging from zero to infinity, one 
recovers the ratio of the  usual leading order cross sections in terms of 
collinear PDFs and FFs. Its agreement with experimental data has been discussed, 
for instance, in Refs.~\cite{Airapetian:2012ki} and~\cite{Signori:2013mda}.     

\section{\label{Results} Results}

As mentioned in the introduction, the most recent analyses of HERMES and COMPASS 
Collaborations provide (unintegrated) multivariate experimental data, presented 
in bins of $\xb = x,\; Q^2,\; z_h =z$ and $P_T$.  The HERMES multiplicities refer 
to identified hadron productions ($\pi^+$, $\pi^-$, $K^+$, $K^-$) off proton and 
deuteron targets, and are presented in $6$ bins of definite $Q^2$ and $\xb$ values, 
each for several different values of $z_h$ and $P_T$, for a total of $2\,660$ 
data points. The selected events cover the kinematical region of $Q^2$ values between 
$1$ and $10$ GeV$^2$ and $0.023 \le \xb \le 0.6$, with a hadronic transverse momentum 
$P_T < 2$ GeV and a fractional energy $z_h$ in the range $0.1 \le z_h \le 0.9$. 

Instead, the COMPASS multiplicities refer to unidentified charged hadron production 
($h^+$ and $h^-$) off a deuteron target ($^6$LiD), and are presented in $23$ bins 
of definite $Q^2$ and $\xb$ values, each for several values of $z_h$ and $P_T$, for 
a total of $18\,624$ data points. The $Q^2$ and $z_h$ regions covered by COMPASS are 
comparable to those explored by the HERMES experiment, while they span a region of 
smaller $\xb$ values, $0.0045 \le \xb \le 0.12$, and cover a wider $P_T$ region 
(reaching lower $P_T$ values).
Moreover, the binning choices are very different and COMPASS statistics is much 
higher than that of HERMES. 

For all these reasons, we consider the two data sets separately and 
perform individual fits.

\subsection{\label{Hermes} Fit of the HERMES multiplicities}

The first step of our analysis consists in using the simple Gaussian 
parameterisation of Eqs.~(\ref{unp-dist}) and (\ref{unp-frag}) and the
expression~(\ref{mult-gaus}), to perform a two parameter fit of the HERMES 
multiplicities $M_n^h(x,\;Q^2,\;z, \;P_T)$. The values of the best fit parameters, 
the Gaussian widths $\avk$ and $\avp$, will fix the TMD distribution and 
fragmentation functions respectively. We do not introduce any overall 
normalisation constant. 

To make sure we work in the region of validity of our simple version of TMD 
factorisation, Eq.~(\ref{mult-gaus}), we further restrict the kinematical 
range explored by the HERMES experiment. In fact, previous studies of the HERMES 
Collaboration~\cite{Airapetian:2012ki} showed that the LO collinear SIDIS cross 
sections (obtained by integration of Eq.~(\ref{mult-gaus}) over $\bfP_T$), 
agree reasonably well with data only in regions of moderate values of $z$. 
The collinear distribution and fragmentation functions which perform best are 
the CTEQ6L PDF set~\cite{Pumplin:2002vw} and the DSS~\cite{deFlorian:2007aj} FFs, 
which we use here. We then consider two possible data selections: $z < 0.7$ and 
$z < 0.6$. Notice that these choices also avoid contaminations from exclusive 
hadronic production processes and large $z$ re-summation 
effects~\cite{Anderle:2012rq}. We also fix the same minimum $Q^2$ as in the 
CTEQ6L analysis, $Q^2 > 1.69$ GeV$^2$, that amounts to excluding the first two 
HERMES $Q^2$ bins.
%
\begin{table}[b]
\caption{$\chi^2$ values of our best fits, following Eqs.~(\ref{mult-gaus}) and 
(\ref{avPT}), of the experimental HERMES measurements of the SIDIS multiplicities 
$M_n^h(\xb, Q^2,z_h, P_T)$ for $\pi^+$ and $\pi^-$ production, off proton and 
deuteron targets. We show the total $\chi^2_{\rm dof}$ and, separately, the
$\chi^2_{\rm point}$ for $\pi^+$ and $\pi^-$ data. CTEQ6 PDFs and DSS FFs are used. 
Notice that the errors quoted for the parameters are statistical errors only, 
and correspond to a $5$\% variation over the total minimum $\chi^2$.
\label{tab:chi-sq-hermes}}
\vspace*{6pt}
\begin{ruledtabular}
\begin{tabular}{cccccc}
\noalign{\vspace{8pt}}
\multicolumn{6}{c}{ \emph{HERMES}} \\
\noalign{\vspace{8pt}}
\cline{1-6}
\noalign{\vspace{8pt}}
  Cuts       & $\chi^2_{\rm dof}$ & n. points  & $[\chi^2_{\rm point}]^{\pi^+}$ &$[\chi^2_{\rm point}]^{\pi^-}$  & Parameters \hspace{10pt} \\
\noalign{\vspace{8pt}}
\cline{1-6}
\noalign{\vspace{8pt}}  
$Q^2 > 1.69\; \textrm{GeV}^2$         &      &     &          &      &  $\langle k_\perp^2 \rangle =  0.57 \pm  0.08\; \textrm{GeV}^2\hspace{10pt}$\\
$0.2 < P_T < 0.9\; \textrm{GeV}$      &1.69  &497  &1.93      & 1.45 &  $\langle p_\perp^2 \rangle =  0.12 \pm  0.01\; \textrm{GeV}^2\hspace{10pt}$\\
$  z < 0.6      $                     &      &     &          &      &   \\  
\noalign{\vspace{5pt}}
\cline{1-6}
\noalign{\vspace{8pt}}  
$Q^2 > 1.69\; \textrm{GeV}^2$        &      &     &         &      &  $\langle k_\perp^2 \rangle =  0.46 \pm  0.09\; \textrm{GeV}^2 \hspace{10pt}$ \\
$0.2 < P_T < 0.9\; \textrm{GeV}$     & 2.62 &576  & 2.56    & 2.68 &  $\langle p_\perp^2 \rangle =  0.13 \pm  0.01\; \textrm{GeV}^2 \hspace{10pt}$\\
$  z < 0.7      $                    &      &     &         &      &   \\
\noalign{\vspace{5pt}}
\end{tabular}
\end{ruledtabular}
\end{table}
%
%
%
Low $P_T$ HERMES data show peculiar deviations from the Gaussian behaviour, 
which instead are not visible in the COMPASS and 
JLab~\cite{Osipenko:2008aa,Mkrtchyan:2007sr} data: for this reason 
we prefer not to consider the lowest $P_T$ bin in order to explore the regions 
which exhibit the same kind of behaviour for all experiments. 
Finally, we apply an additional cut on large $P_T$, requiring $P_T < 0.9$ GeV, 
as multiplicities at 
large $P_T$ values fall in the domain of the onset of collinear perturbative 
QCD~\cite{Anselmino:2006rv}. In the considered $Q^2$ range, this implies 
$P_T/Q < 0.7$. Notice that recent analyses of the same experimental 
data~\cite{Sun:2013hua, Signori:2013mda} have adopted similar cuts. 

Summarising, we limit the analysis of HERMES SIDIS data to the kinematical 
regions defined by:
\bea
&&\qquad z < 0.7 \qquad Q^2 > 1.69 \; \textrm{GeV}^2  
\qquad 0.2 < P_T < 0.9 \; \textrm{GeV} \label{kin-cuts1}\\
&& \qquad z < 0.6 \qquad Q^2 > 1.69 \; \textrm{GeV}^2  
\qquad 0.2 < P_T < 0.9 \; \textrm{GeV} 
\;. \label{kin-cuts2}
\eea
Moreover, in our fit we do not include 
the kaon production data points; in fact, the precision and accuracy 
of the kaon data sample, at present, do not help in constraining the 
values of the free parameters. When taken into account, the kaon data have 
little or no impact on the fit and are compatible with the assumption of the 
same Gaussian width as for pion production. This will be explicitly shown
below by computing, using the parameters extracted from pion data, the kaon
multiplicities and comparing them with the HERMES results. 

The above selections reduce the number of fitted HERMES data points 
to either $576$ for $z<0.7$, or $497$ for $z<0.6$.

The details of the fits are presented in Table~\ref{tab:chi-sq-hermes}, where 
we show the $\chi^2$ per degree of freedom ($\chi^2_{\rm dof}$), the $\chi^2$ 
per number of points ($\chi^2_{\rm point}$) for $\pi^+$ and $\pi^-$ production 
and the resulting values of the two free parameters of the fit, $\avk$ and $\avp$, 
with some statistical errors, as explained below. It is worth noticing again that 
we do not have to use any overall normalisation constant as an extra free 
parameter; our computations agree well in magnitude with the experimental 
multiplicities, which are normalised to the collinear DIS cross section. 

Before drawing hasty conclusions on the numerical values of the parameters, 
some comments might be helpful. 
\begin{itemize}
\item
Our lowest value of $\chi^2_{\rm dof}$ is obtained by 
using the kinematical cuts of Eq.~(\ref{kin-cuts2}) with $z < 0.6$,
$\chi^2_{\rm dof} = 1.69$ for a total of $497$ fitted pion data points. 
The corresponding widths of the Gaussians representing the $\kt$ and $\pp$ 
dependences of the distribution and fragmentation functions, are:
\be
\langle \kt^2 \rangle = 0.57 \pm 0.08\; \textrm{GeV}^2 \;, \hspace{1cm}  
\langle \pp^2 \rangle = 0.12 \pm 0.01\; \textrm{GeV}^2 \;.
\label{hermes-par}
\ee
However, if we relax the cut in $z$ to $z<0.7$, Eq.~(\ref{kin-cuts1}), then 
the total $\chi^2$ of the fit becomes larger, $\chi^2_{\rm dof}=2.62$, and the 
value of the extracted $\avk$ Gaussian width significantly decreases while that 
of $\avp$ increases, as shown in the second row of Table~\ref{tab:chi-sq-hermes}. 
This large value of $\chi^2$ reflects the fact that, at large values of $z$, 
$\avPT$ deviates from the assigned linear behaviour in $z^2$. 
Morever, as we already pointed out, the large $z$ region suffers from our lack 
of knowledge on the collinear fragmentation functions.
\item
The errors quoted for the free parameters of our fit are obtained from a 
$\Delta \chi ^2$ corresponding to a $5$\% variation over the total minimum 
$\chi^2$: following Ref.~\cite{Epele:2012vg}, we relax the 
usual choice of $\Delta \chi ^2 =1$, corresponding to a purely statistical 
error, in order to include in the quoted errors other, major sources 
of uncertainty in our fit, which mainly originate from the inaccuracy in the 
determination of the fragmentation functions. We have checked that, indeed, 
other choices of collinear PDFs and FFs lead to such uncertainties.  
Moreover, in reading the errors, one should keep in mind that the 
parameters are strongly correlated.     
\end{itemize}

\begin{figure}
\begin{center}
\includegraphics[width=0.6\textwidth, angle=-90]{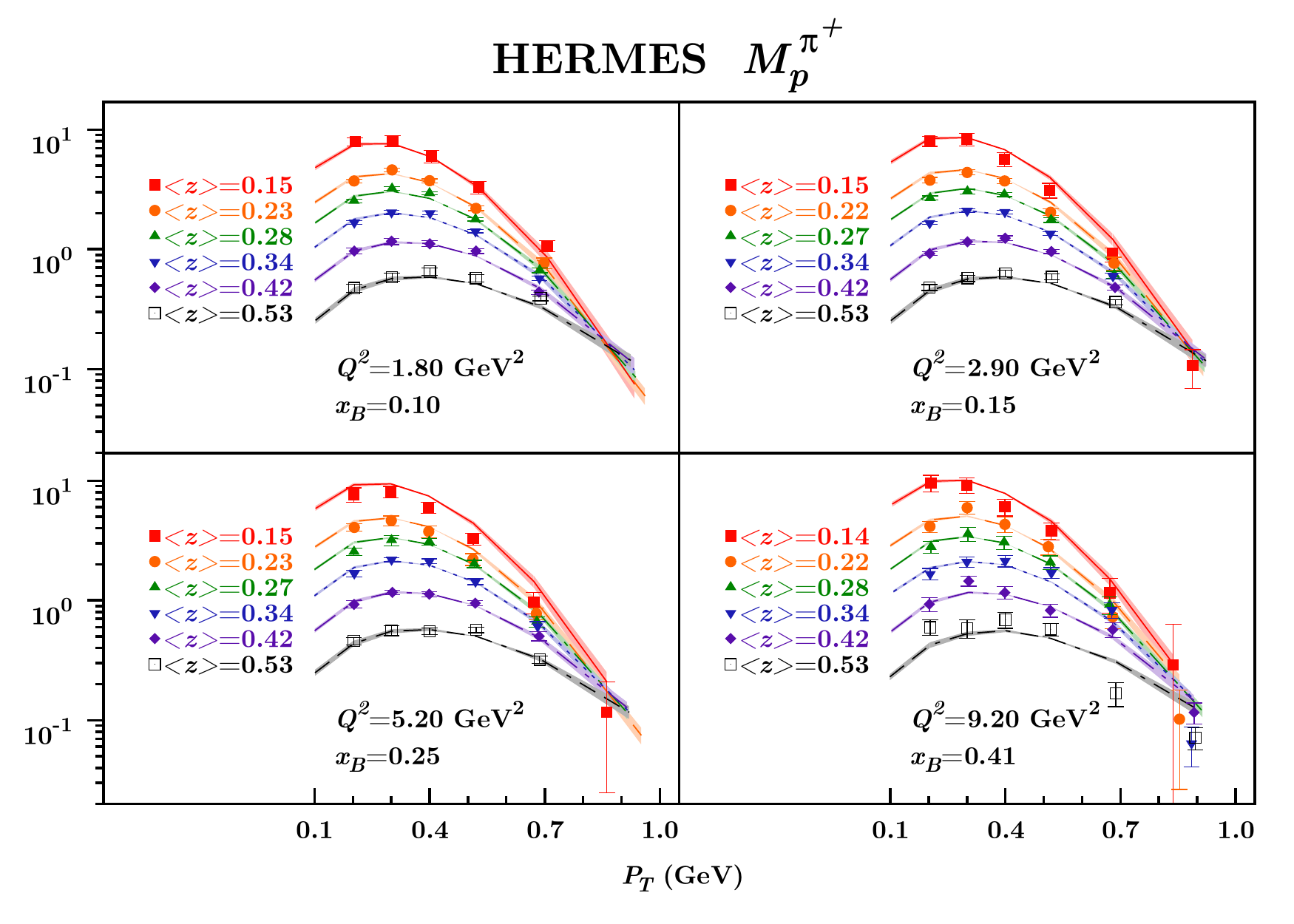}
\caption{\label{fig:hermes-p-pi+}
The multiplicities $M_p^{\pi^+}$ obtained from Eqs.~(\ref{mult-gaus}) and 
(\ref{avPT}), with the parameters of Eq.~(\ref{hermes-par}), are compared with 
HERMES measurements for $\pi^+$ SIDIS production off a proton 
target~\cite{Airapetian:2012ki}. The shaded uncertainty bands correspond
to a $5$\% variation of the total $\chi^2$.}
%
\includegraphics[width=0.6\textwidth, angle=-90]{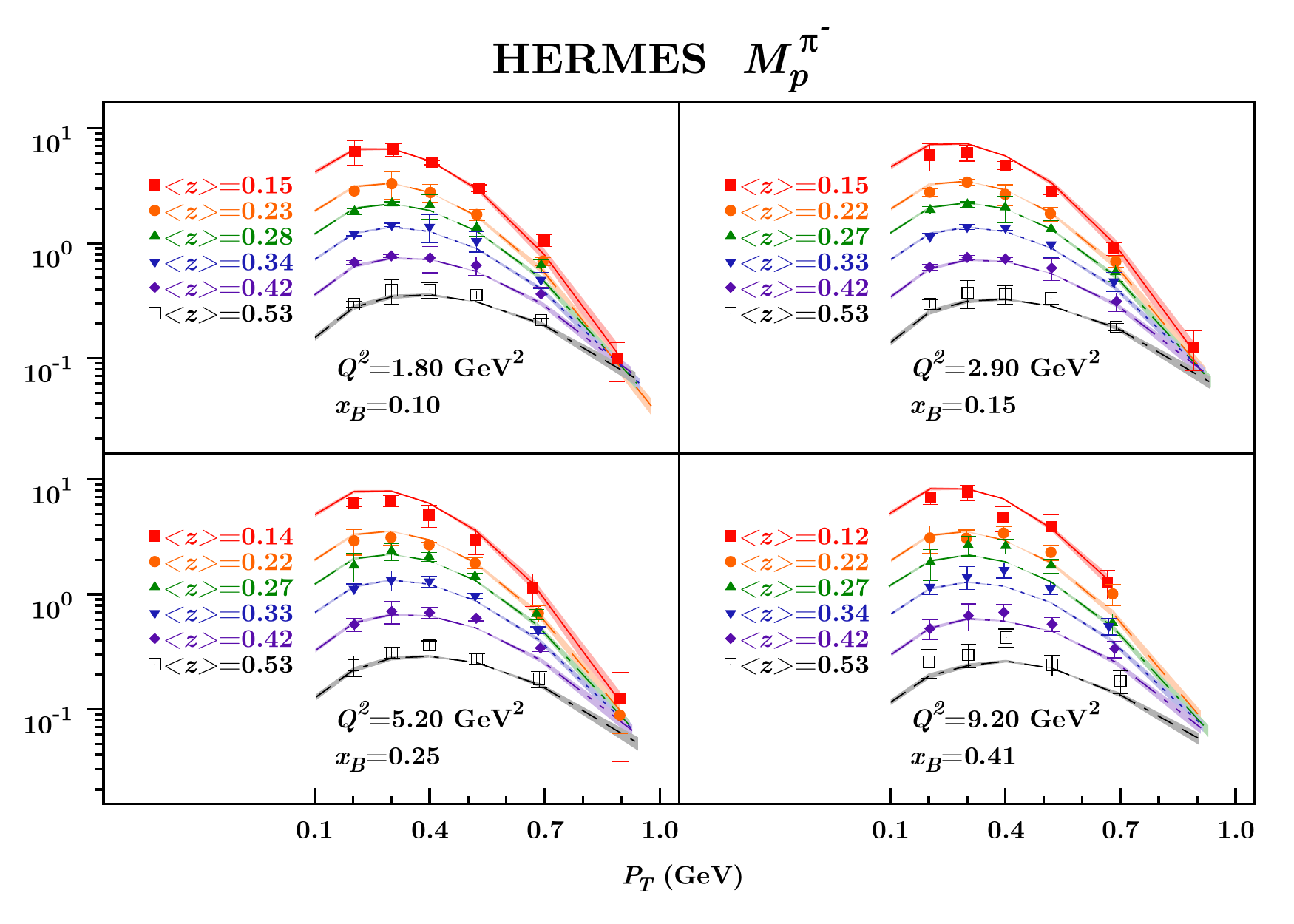}
\caption{\label{fig:hermes-p-pi-}
The multiplicities $M_p^{\pi^-}$ obtained from Eqs.~(\ref{mult-gaus}) and 
(\ref{avPT}), with the parameters of Eq.~(\ref{hermes-par}), are compared with 
HERMES measurements for $\pi^-$ SIDIS production off a proton 
target~\cite{Airapetian:2012ki}. The shaded uncertainty bands correspond
to a $5$\% variation of the total $\chi^2$.}
\end{center}
\end{figure}

\begin{figure}
\includegraphics[width=0.6\textwidth, angle=-90]{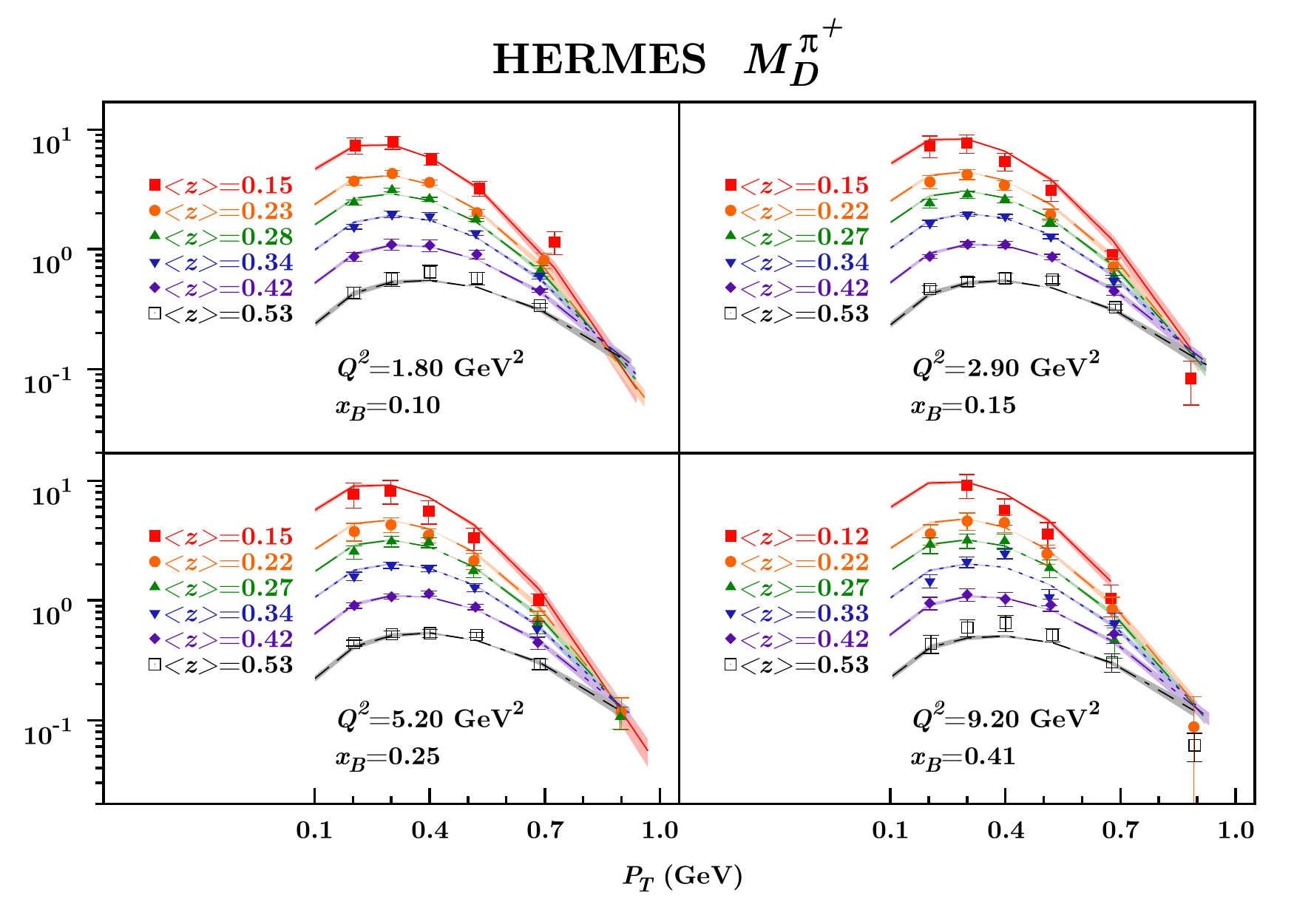}
\caption{\label{fig:hermes-D-pi+}
The multiplicities $M_D^{\pi^+}$ obtained from Eqs.~(\ref{mult-gaus}) and 
(\ref{avPT}), with the parameters of Eq.~(\ref{hermes-par}), are compared with 
HERMES measurements for $\pi^+$ SIDIS production off a deuteron 
target~\cite{Airapetian:2012ki}. The shaded uncertainty bands correspond
to a $5$\% variation of the total $\chi^2$.}
\includegraphics[width=0.6\textwidth, angle=-90]{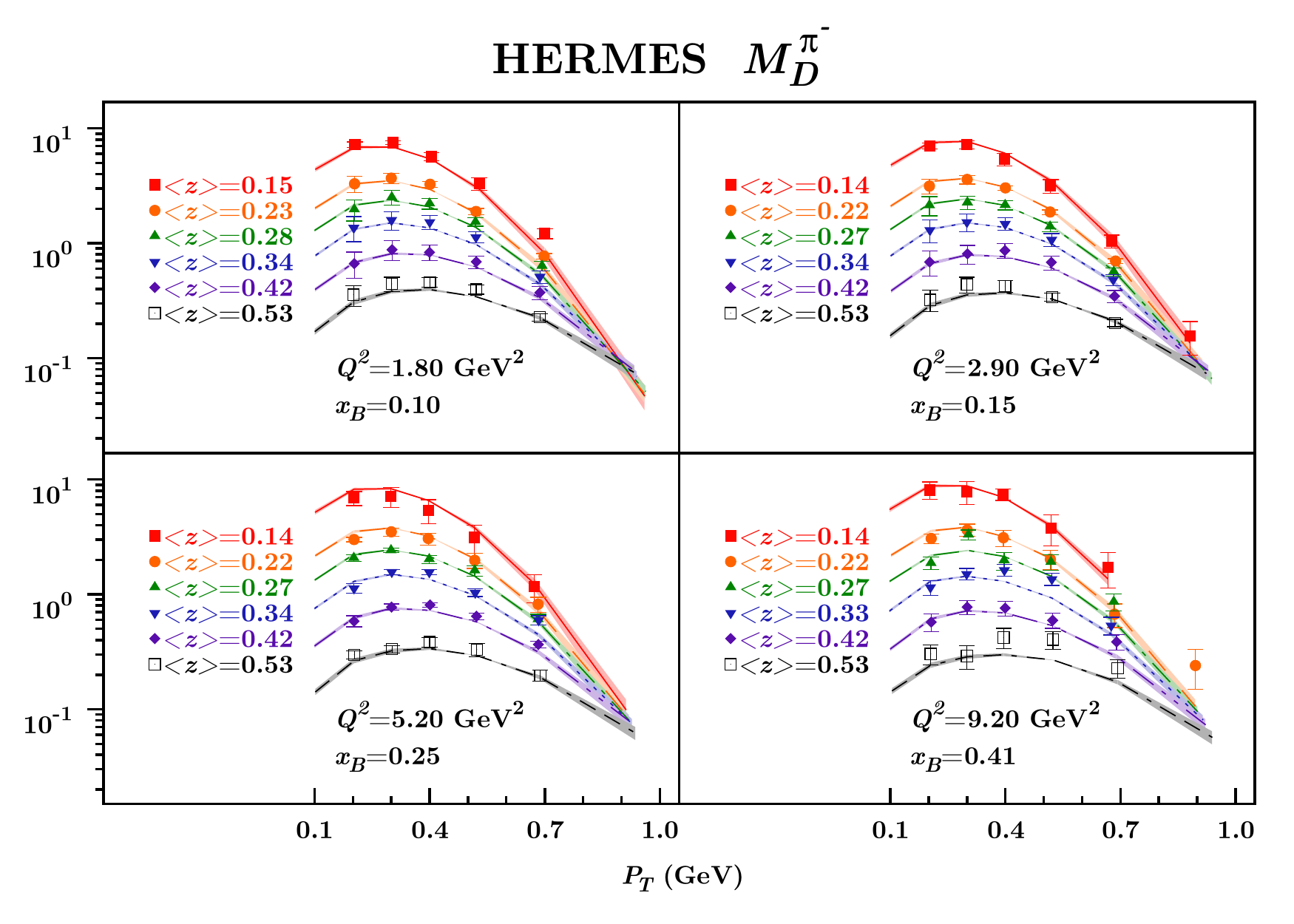}
\caption{\label{fig:hermes-D-pi-}
The multiplicities $M_D^{\pi^-}$ obtained from Eqs.~(\ref{mult-gaus}) and 
(\ref{avPT}), with the parameters of Eq.~(\ref{hermes-par}), are compared with 
HERMES measurements for $\pi^-$ SIDIS production off a deuteron 
target~\cite{Airapetian:2012ki}. The shaded uncertainty bands correspond
to a $5$\% variation of the total $\chi^2$.}
\end{figure}

The multiplicities obtained from our best fit parameters, with the 
kinematical cuts of Eq.~(\ref{kin-cuts2}), are compared with the 
HERMES measurements off a proton target in Figs.~\ref{fig:hermes-p-pi+} 
and~\ref{fig:hermes-p-pi-} and off a deuteron target in 
Figs.~\ref{fig:hermes-D-pi+} and~\ref{fig:hermes-D-pi-}, separately for positive 
and negative pions. The shaded uncertainty bands are computed according to 
Ref.~\cite{Barone:2006xj}. 

\begin{figure}
\includegraphics[width=0.58\textwidth, angle=-90]{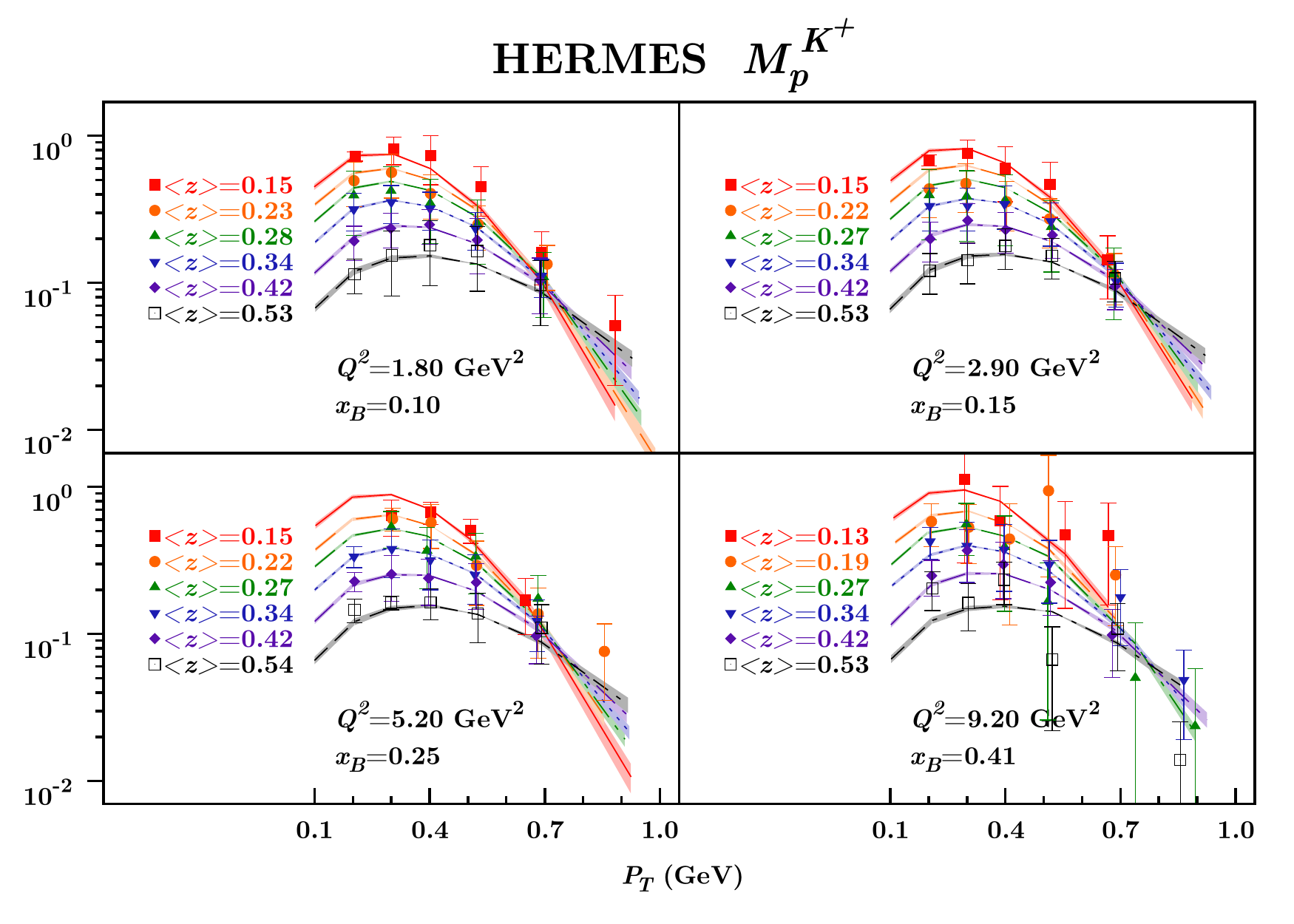}
\caption{\label{fig:hermes-p-K+}
The multiplicities $M_p^{K^+}$ obtained from Eqs.~(\ref{mult-gaus}) and 
(\ref{avPT}), with the parameters of Eq.~(\ref{hermes-par}), are compared with 
HERMES measurements for $K^+$ SIDIS production off a proton 
target~\cite{Airapetian:2012ki}. Notice that these data are not included in the fit; 
the shaded uncertainty bands correspond
to a $5$\% variation of the total $\chi^2$ obtained by fitting pion data only.}
\includegraphics[width=0.58\textwidth, angle=-90]{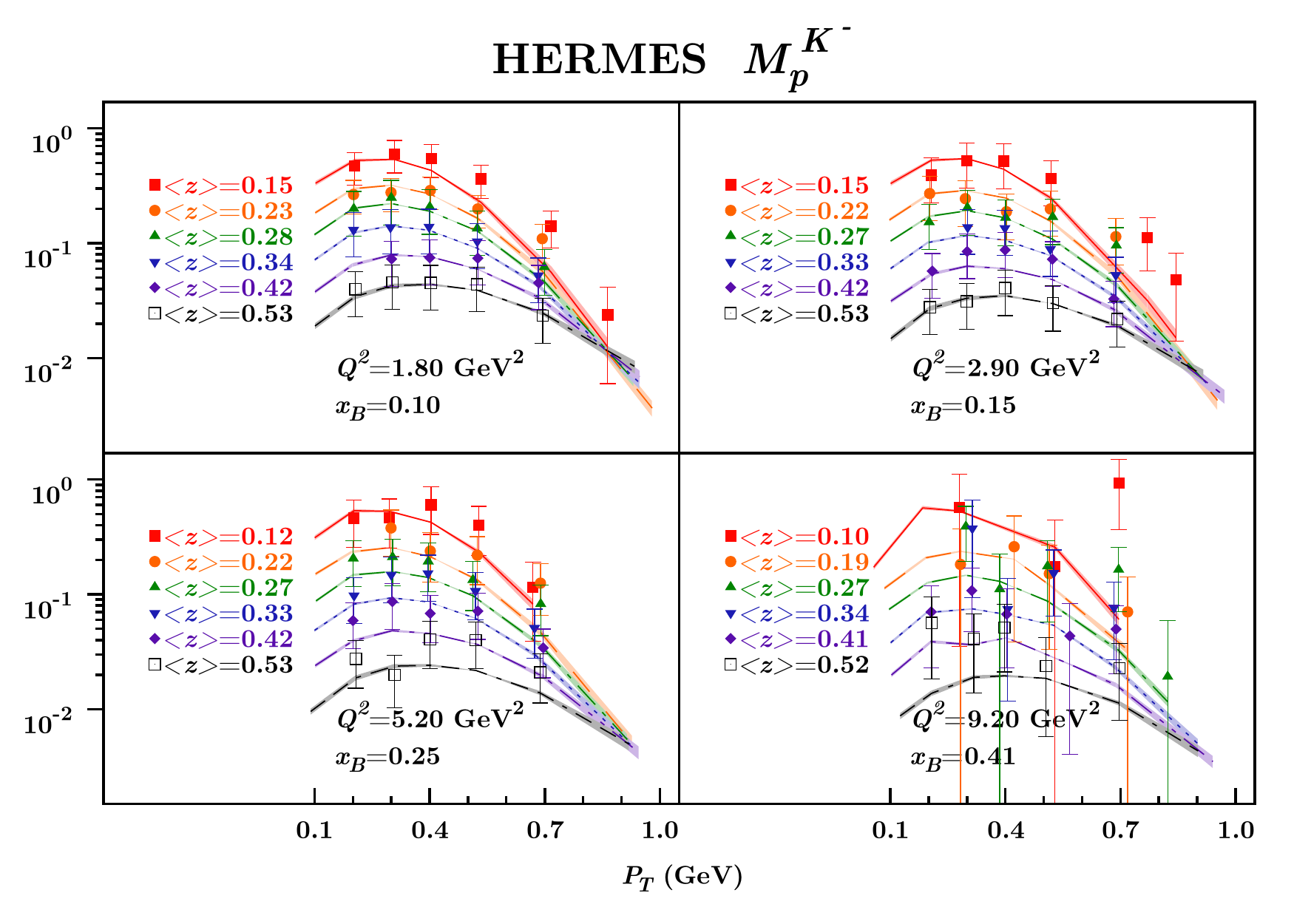}
\caption{\label{fig:hermes-p-K-}
The multiplicities $M_p^{K^-}$ obtained from Eqs.~(\ref{mult-gaus}) and 
(\ref{avPT}), with the parameters of Eq.~(\ref{hermes-par}), are compared with 
HERMES measurements for $K^-$ SIDIS production off a proton 
target~\cite{Airapetian:2012ki}. Notice that these data are not included in the fit; 
the shaded uncertainty bands correspond
to a $5$\% variation of the total $\chi^2$ obtained by fitting pion data only.}
 \end{figure}

\begin{figure}
\includegraphics[width=0.58\textwidth, angle=-90]{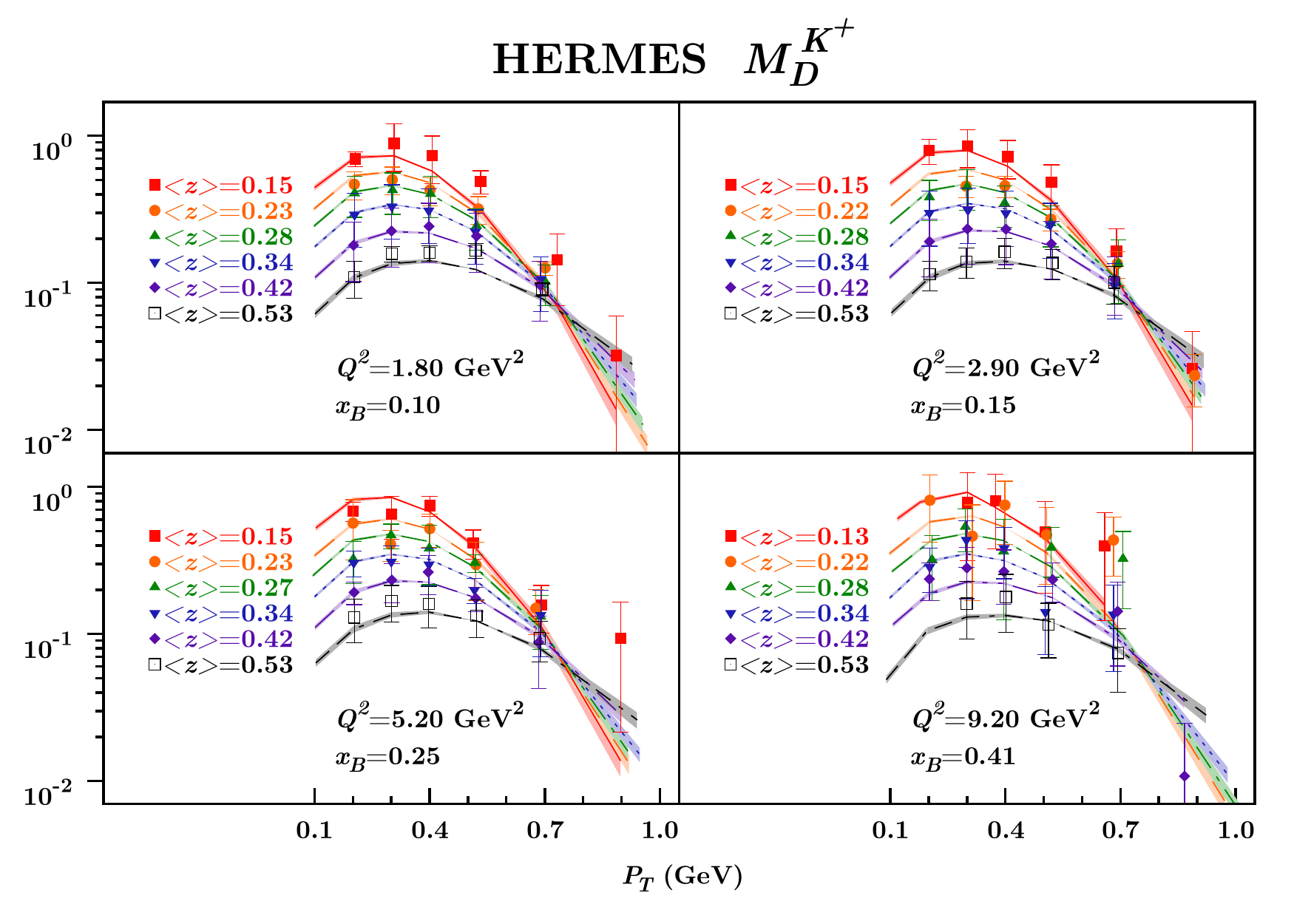}
\caption{\label{fig:hermes-D-K+}
The multiplicities $M_D^{K^+}$ obtained from Eqs.~(\ref{mult-gaus}) and 
(\ref{avPT}), with the parameters of Eq.~(\ref{hermes-par}), are compared with 
HERMES measurements for $K^+$ SIDIS production off a deuteron 
target~\cite{Airapetian:2012ki}. Notice that these data are not included in the fit; 
the shaded uncertainty bands correspond
to a $5$\% variation of the total $\chi^2$ 
obtained by fitting pion data only.}
\includegraphics[width=0.58\textwidth, angle=-90]{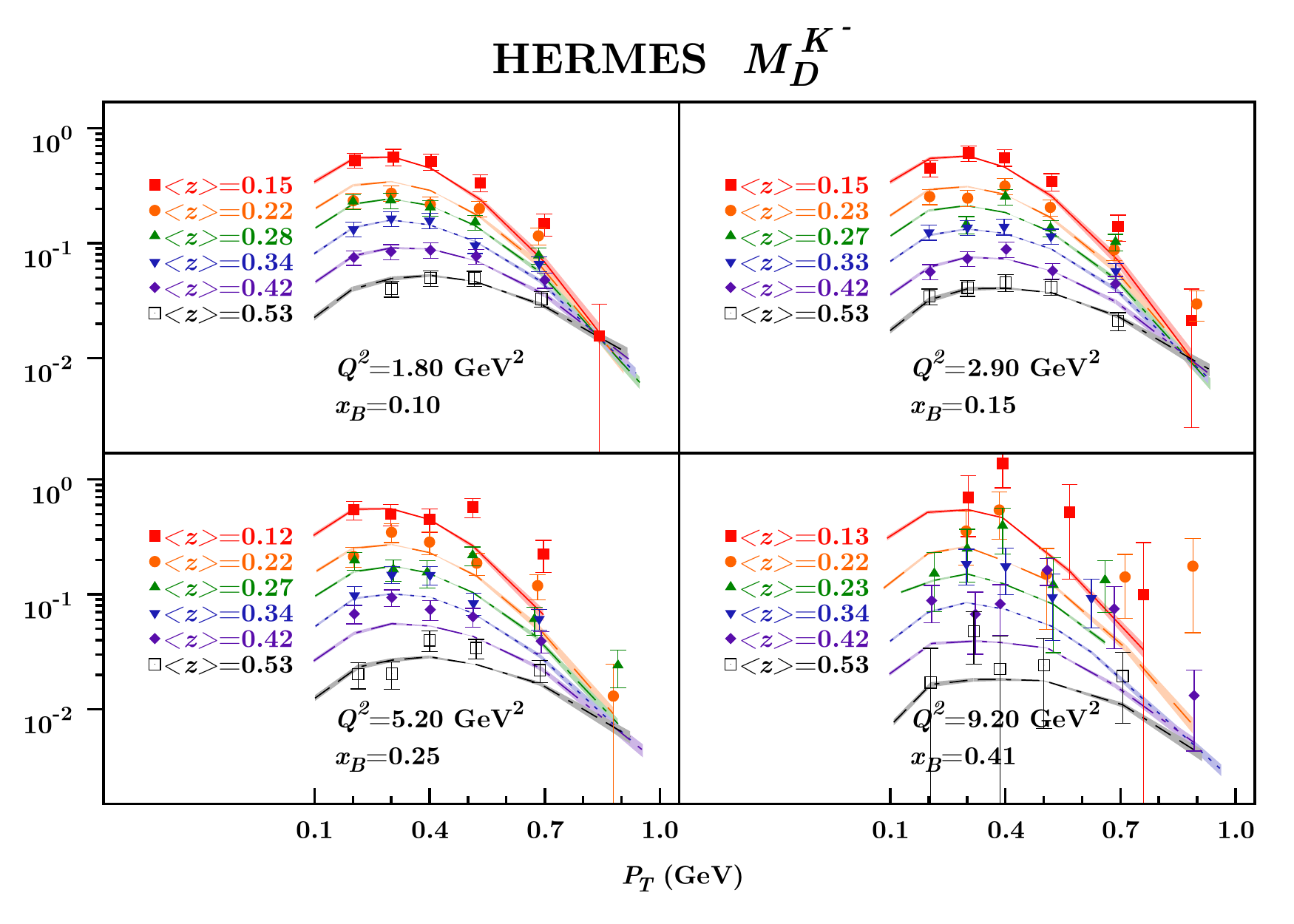}
\caption{\label{fig:hermes-D-K-}
The multiplicities $M_D^{K^-}$ obtained from Eqs.~(\ref{mult-gaus}) and 
(\ref{avPT}), with the parameters of Eq.~(\ref{hermes-par}), are compared with 
HERMES measurements for $K^-$ SIDIS production off a deuteron 
target~\cite{Airapetian:2012ki}. Notice that these data are not included in the fit; 
the shaded uncertainty bands correspond
to a $5$\% variation of the total $\chi^2$ obtained by fitting pion data only.}
\end{figure}

We have also performed a series of tests to study the effect of kaon data
on the extraction. While the optimal parameters do not significantly change 
when including these data in the fit, the value of $\chi^2_{\rm dof}$ reduces 
from $1.69$ to $1.25$, which could naively be interpreted as an improvement in 
the quality of the fit. However, this is just the result of the large error 
bands in the kaon subset. In fact, a fit of the kaon data alone would yield 
$\chi^2_{\rm dof} = 0.64$, which signals that the errors on these 
measurements are too large to allow a reliable extraction of the free 
parameters. This is shown very clearly 
in Figs.~\ref{fig:hermes-p-K+}--\ref{fig:hermes-D-K-} where the kaon 
multiplicities, computed according to Eqs.~(\ref{mult-gaus}) and (\ref{avPT}) 
with the parameters of Eq.~(\ref{hermes-par}) -- extracted from the HERMES 
measurements of pion production only -- are compared with the HERMES data.

A careful look at the plots in Figs.~\ref{fig:hermes-p-pi+}--\ref{fig:hermes-D-pi-} 
shows that the description of the HERMES measurements is indeed satisfactory: 
the Gaussian parameterisation embeds the crucial features of the data, both in 
shape and size, over a broad kinematical range. The resulting value of 
$\chi^2_{\rm dof}$, still a bit sizeable, is somehow expected, given the 
uncertainties on the collinear fragmentation functions: as stated before, the 
HERMES analysis~\cite{Airapetian:2012ki} showed that the agreement, for the 
integrated multiplicities, between SIDIS data and collinear LO theoretical 
computations is not perfect, and that the currently available 
fragmentation function sets still need further refinements, especially at 
large $z$, and for $\pi^-$ production. In fact, including larger values of $z$ 
in the fit sizeably increases the total $\chi^2$, as shown in the second line of 
Table~\ref{tab:chi-sq-hermes}. 

As the HERMES Montecarlo event generator, as well as many phenomenological 
models, propose a possible dependence of $\avp$ on $z$, we have also attempted 
a fit with a $z$-dependent $\avp = N\,z^a\,(1-z)^b$. However, it turns out that 
this parameterisation cannot be seriously tested by the data selection we have 
used for our reference fit; in fact, with the cuts of Eqs.~(\ref{kin-cuts1})
and (\ref{kin-cuts2}), and in particular for the $z < 0.6$ range, 
it is quite hard for the best fitting procedure to find a proper convergence. 
Consequently, one obtains $a$ and $b$ parameters affected by huge statistical 
errors; this large uncertainties include the zero value and make the resulting 
parameters hardly significant. Moreover, the total $\chi^2_{\rm dof}$ improves 
only marginally, down from 1.69 to 1.63.   
%
%
%
%
%
%

To try and obtain more stringent constraints on this unknown $z$-dependence, we 
included also all the large-$z$ bins, 
thus increasing the number of fitted data 
points from $497$ to $657$: in this case the minimisation procedure converges 
easily and the sizes of the errors on $a$ and $b$ become acceptable, excluding 
the zero values. 
However, the $\chi^2$ we obtain, $\chi^2_{\rm dof} = 2.92$, becomes large, 
reflecting the large uncertainty in the collinear fragmentation functions at 
large $z$, and does not allow a reliable interpretation of the physical meaning 
of the extracted parameters. For this reason, we believe that these measurements 
are unable to fix the precise $z$-dependence of the Gaussian width $\avp$.   

A study of the possible dependence of $\avk$ on $x$, similar to that of 
Ref.~\cite{Signori:2013mda}, shows no significant effect for the HERMES data. 
Notice that some $x$ dependence of $\avk$ will be further considered in 
Section IV, in the framework of the TMD evolution. 

Similarly, no significant improvement is obtained by allowing flavour dependent 
$\avk$ Gaussian widths. However, it is interesting to notice that, within a 
parameterisation in which we allow for two different free parameters for the 
favoured and disfavoured fragmentation Gaussian widths, HERMES data seem to 
prefer a configuration in which the favoured $\avp_{\rm fav}$ is smaller than 
the disfavoured $\avp_{\rm disf}$ by about 15\%: we obtain $\avp_{\rm fav} = 
0.12 \pm 0.01$ and $\avp_{\rm disf} = 0.14 \pm 0.01$, together with 
$\avk = 0.59 \pm 0.08$, corresponding to a total $\chi^2_{\rm dof} = 1.60$. 
These results are in agreement with those obtained in Ref.~\cite{Signori:2013mda}. 

We can therefore conclude that the HERMES multiplicity measurements, in the 
selected kinematical regions, can satisfactorily be described by modelling the 
intrinsic transverse momentum dependence with a very simple Gaussian shape with 
a constant width. Only a very slight indication for a flavour dependence of the 
fragmentation width is found, with the disfavoured Gaussian slightly wider than 
the favoured one. We also notice that, presently, the HERMES data do not show 
any sign of $Q^2$ evolution. The scale dependence will be discussed in 
Section IV. 

\subsection{\label{Compass} Fit of the COMPASS multiplicities}

We now consider the COMPASS SIDIS multiplicities of Ref.~\cite{Adolph:2013stb} 
and best fit them separately.

We apply the same cuts as those used for the HERMES set of data, 
Eq.~(\ref{kin-cuts1}) and (\ref{kin-cuts2}), with the same motivations, which 
reduce the number of fitted data points to $6\,284$ for $z<0.7$ and $5\,385$ for 
$z<0.6$. Our resulting $\chi^2_{\rm dof}$ values are presented in 
Table~\ref{tab:chi-sq-compass} and turn out to be much larger 
($\chi^2_{\rm dof} = 9.81$ and $8.54$) than those obtained by fitting the HERMES 
multiplicities.

Taken at face value the parameters extracted from our best fit (first line of 
Table~\ref{tab:chi-sq-compass}), 
\be
\avk =  0.61 \pm  0.20\; \textrm{GeV}^2 \qquad
\avp =  0.19 \pm  0.02\; \textrm{GeV}^2 \,,
\label{compass-par}
\ee
indicate that COMPASS data lead to values of $\avk$ consistent, within errors, 
with those obtained in the HERMES fit, Eq.~(\ref{hermes-par}), while $\avp$ seems 
to be slightly larger (considering the tiny error in the determination of this 
parameter). As in the HERMES fit, by widening the range in $z$ the total $\chi^2$ 
increases, although in this case the values of the extracted parameters are 
much more stable.

\begin{figure}
\includegraphics[width=0.8\textwidth, angle=-90]{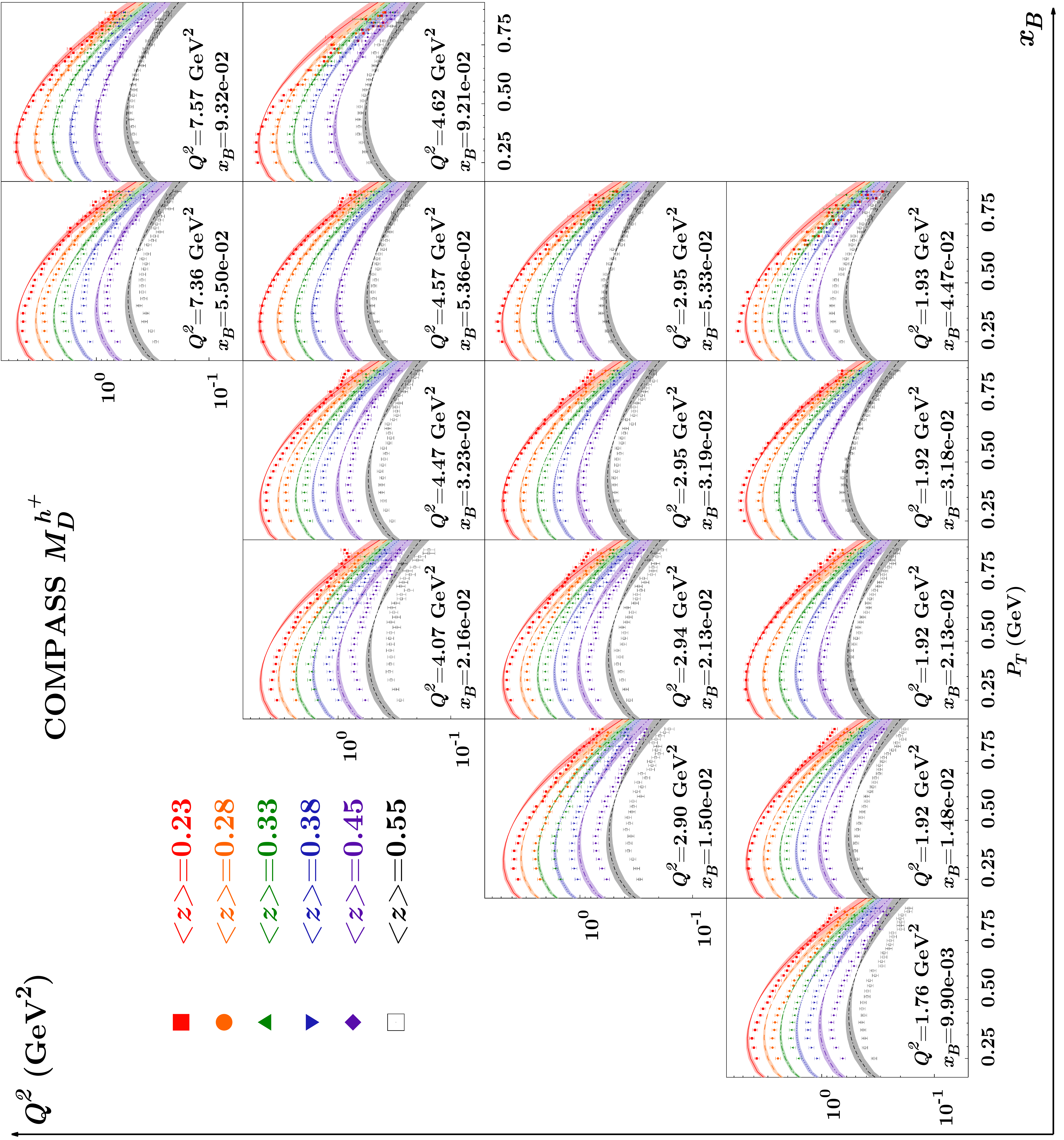}
\caption{\label{fig:compass-D-h+}
The multiplicities $M_D^{h^+}$ obtained from Eqs.~(\ref{mult-gaus}) and 
(\ref{avPT}), with the parameters of Eq.~(\ref{compass-par}), are compared with COMPASS measurements 
for $h^+$ SIDIS production off a deuteron 
target~\cite{Adolph:2013stb}. The shaded uncertainty bands correspond
to a $5$\% variation of the total $\chi^2$.}
\end{figure}

\begin{figure}
\includegraphics[width=0.8\textwidth, angle=-90]{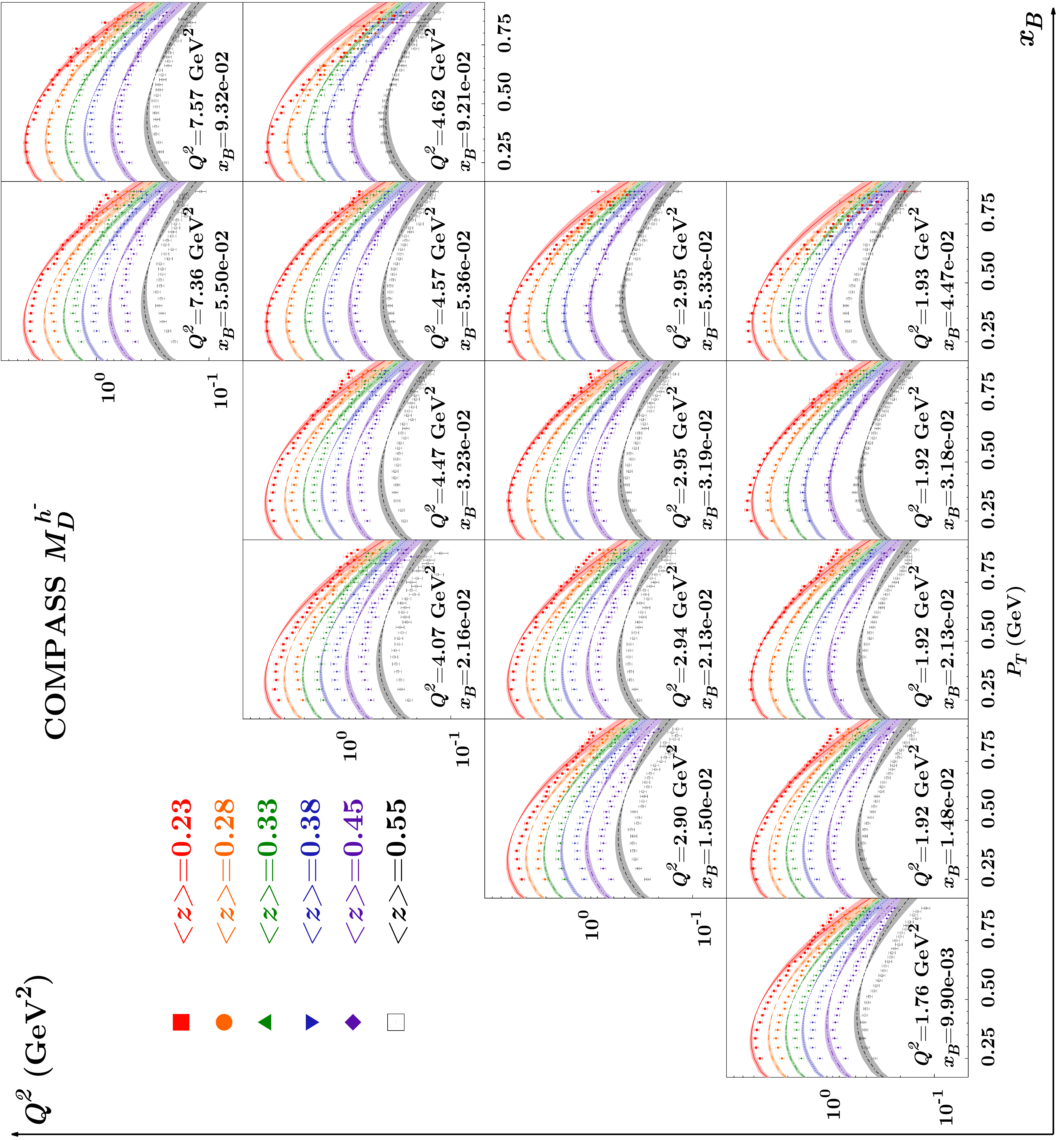}
\caption{\label{fig:compass-D-h-}
The multiplicities $M_D^{h^-}$ obtained from Eqs.~(\ref{mult-gaus}) and 
(\ref{avPT}), with the parameters of Eq.~(\ref{compass-par}), are compared with COMPASS measurements 
for $h^-$ SIDIS production off a deuteron 
target~\cite{Adolph:2013stb}. The shaded uncertainty bands correspond
to a $5$\% variation of the total $\chi^2$.}
\end{figure}

Our best fit curves ($z < 0.6$), for the COMPASS multiplicities 
$M_D^{h^\pm}(\xb, Q^2,z_h, P_T)$, are shown in Figs.~\ref{fig:compass-D-h+} 
and \ref{fig:compass-D-h-}, where it is clear that the quality of such fits 
appear to be rather poor. 

Attempts of improving the description of the COMPASS data assuming more elaborated 
modelings of the Gaussian widths, with $x$ and $z$ dependence, do not help.  
We have also allowed for flavour dependent values of $\avk$ (valence and sea) 
associated to two (favoured and disfavoured) fragmentation Gaussian widths $\avp$. 
However, even by adding a remarkably large number of free parameters, the 
gain in $\chi^2$ is not satisfactory and the description of the data improves only marginally.

By carefully inspecting Figs.~\ref{fig:compass-D-h+} and \ref{fig:compass-D-h-}, 
and the $\chi^2$ contributions of each individual bin, we realised that major 
improvements of our description of COMPASS data cannot actually be achieved by 
modifying the Gaussian widths, nor by making it more flexible. Our 
simple Gaussian model can actually reproduce the shape of the data, 
even over a large kinematical range; rather, the difficulties of the fit seem to 
reside in the {\it normalisation} of some of the bins, in particular those corresponding to large values of $y$. 

\begin{table}[ht]
\caption{$\chi^2$ values of our best fits, following Eqs.~(\ref{mult-gaus}) and 
(\ref{avPT}), of the experimental COMPASS measurements of the SIDIS multiplicities 
$M_n^h(\xb, Q^2,z_h, P_T)$ for $h^+$ and $h^-$ production, off a 
deuteron target. We show the total $\chi^2_{\rm dof}$ and, separately, the
$\chi^2_{\rm dof}$ for $h^+$ and $h^-$ data. CTEQ6 PDFs and DSS FFs are used. 
Notice that the errors quoted for the parameters are statistical errors only, 
and correspond to a $5$\% variation over the total minimum $\chi^2$.
The two lowest rows of numerical results are obtained allowing for a $y$-dependent
extra normalisation factor, Eq.~(\ref{y-dep}).   
\label{tab:chi-sq-compass}}
\vspace*{6pt}
\begin{ruledtabular}
\begin{tabular}{cccccc}
\noalign{\vspace{8pt}}
\multicolumn{6}{c}{ \emph{COMPASS}} \\
\noalign{\vspace{8pt}}
\cline{1-6}
\noalign{\vspace{8pt}}
  Cuts       & $\chi^2_{\rm dof}$ & n. points  & $[\chi^2_{\rm point}]^{h^+}$ &
  $[\chi^2_{\rm point}]^{h^-}$  & Parameters\hspace{10pt}   \\
\noalign{\vspace{5pt}}
\cline{1-6}
\noalign{\vspace{5pt}}  
$Q^2 > 1.69\; \textrm{GeV}^2$         &      &     &          &      & $\langle k_\perp ^2 \rangle =  0.61 \pm  0.20\; \textrm{GeV}^2\hspace{10pt}$ \\
$0.2 < P_T < 0.9\; \textrm{GeV}$      &8.54  &5385 & 8.94     & 8.15 & $\langle p_\perp^2      \rangle =  0.19 \pm  0.02\; \textrm{GeV}^2\hspace{10pt}$ \\
$  z < 0.6     $                      &      &     &          &      &  \\
\noalign{\vspace{5pt}}
\cline{1-6}
\noalign{\vspace{5pt}}  
$Q^2 > 1.69\; \textrm{GeV}^2$        &      &     &         &      &  $\langle k_\perp ^2 \rangle =  0.57 \pm  0.21\; \textrm{GeV}^2 \hspace{10pt}$\\
$0.2 < P_T < 0.9\; \textrm{GeV}$     & 9.81 &6284  & 10.37    & 9.25 & $\langle p_\perp^2      \rangle =  0.19 \pm  0.02\; \textrm{GeV}^2 \hspace{10pt}$  \\
$  z < 0.7      $                    &      &     &         &      &   \\
\noalign{\vspace{5pt}}
\cline{1-6}
\noalign{\vspace{5pt}}  
$Q^2 > 1.69\; \textrm{GeV}^2$         &      &     &          &      &  $\langle k_\perp ^2 \rangle =  0.60 \pm  0.14\; \textrm{GeV}^2\hspace{10pt}$ \\
$0.2 < P_T < 0.9\; \textrm{GeV}$      &3.42  &5385 &  3.25    & 3.60 &  $\langle p_\perp^2      \rangle =  0.20 \pm  0.02\; \textrm{GeV}^2\hspace{10pt}$ \\
$  z < 0.6    $                       &      &     &          &      &   $A=1.06\pm0.06$ \\  
$N_y= A + B\,y $                      &      &     &          &      &   $B=-0.43\pm0.14$                                                      \\
\noalign{\vspace{5pt}}
\cline{1-6}
\cline{1-6}
\noalign{\vspace{5pt}}                                                        
$Q^2 > 1.69\; \textrm{GeV}^2$         &      &     &          &      &  $\langle k_\perp ^2 \rangle =  0.52 \pm  0.14\; \textrm{GeV}^2\hspace{10pt}$ \\
$0.2 < P_T < 0.9\; \textrm{GeV}$      &3.79  &6284 &  3.63    & 3.96 &  $\langle p_\perp^2      \rangle =  0.21 \pm  0.02\; \textrm{GeV}^2\hspace{10pt}$ \\
$  z < 0.7    $                       &      &     &          &      &   $A=1.06\pm0.07$ \\  
$N_y= A + B\,y $                      &      &     &          &      &   $B=-0.46\pm0.15$                                                      \\
\noalign{\vspace{5pt}}
\end{tabular}
\end{ruledtabular}
\end{table}
%
%
%
%
%
\begin{figure}[ht]
\includegraphics[width=0.4\textwidth, angle=-90]{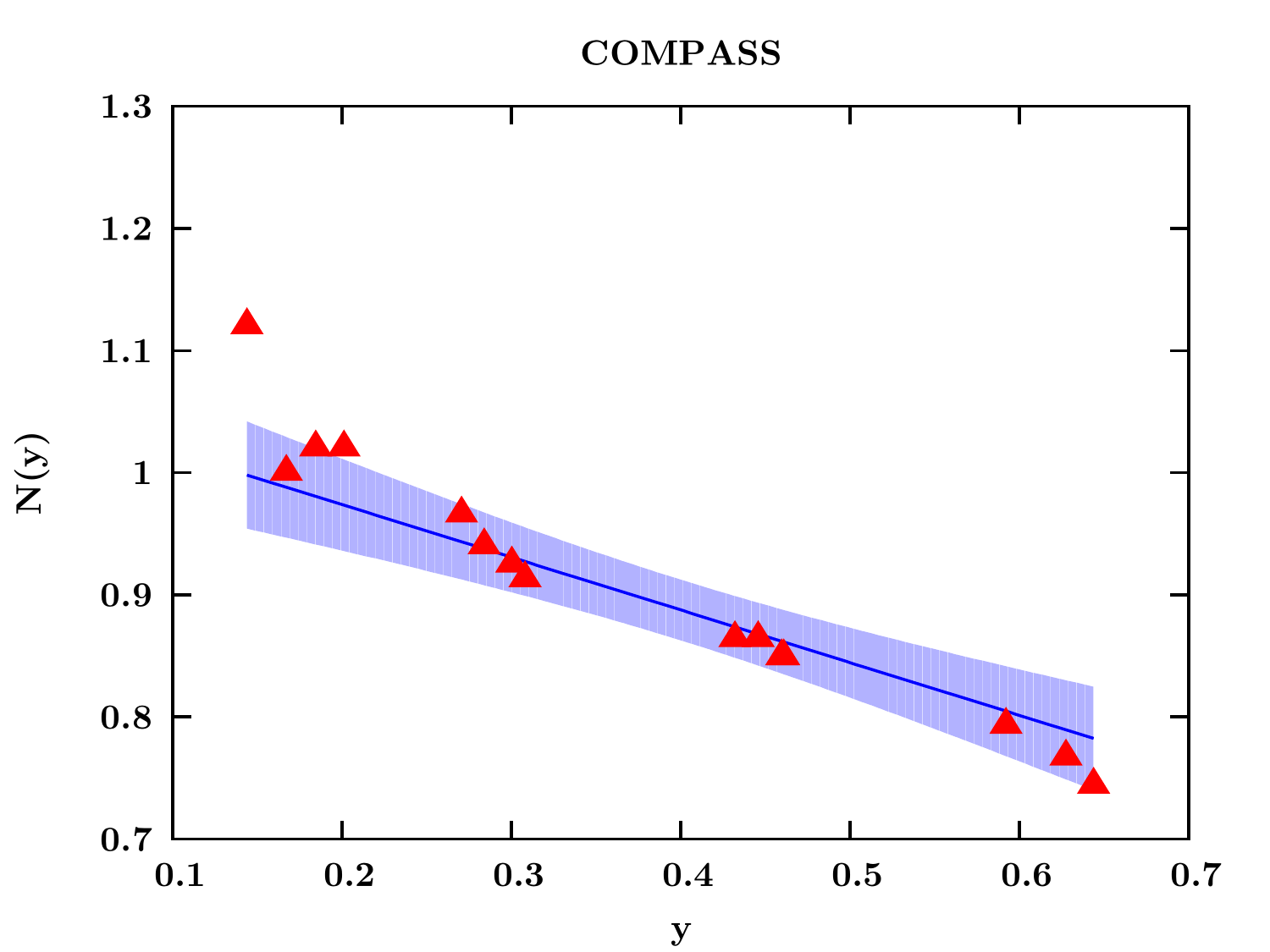}
\caption{\label{fig:N-y}
The $N$ normalisation coefficients, obtained by fitting separately each individual 
($\xb, \, Q^2$) bin, as a function of the kinematical variable $y$. The line is 
given by Eq.~(\ref{y-dep}) with $A = 1.06 \pm 0.06$ and $B = -0.43 \pm 0.14$.}
\end{figure}
%

To analyse this effect more closely, we performed the following study: 
we fitted the same sets of COMPASS data, allowing different normalisation 
constants $N$ for each individual ($\xb,\,Q^2$) bin ({\it i.e.} for each individual 
panel of Figs.~\ref{fig:compass-D-h+} and \ref{fig:compass-D-h-}) and we plotted 
all the obtained values versus $\xb$, $Q^2$ and $y$. Although no particular 
$\xb$ or $Q^2$  trend could be detected, it was immediately evident that the 
obtained $N$ coefficients showed a linear dependence on the kinematical variable 
$y$, as shown in Fig.~\ref{fig:N-y}. This suggested that a better description 
of COMPASS multiplicities could possibly be achieved by assuming a $y$-dependent 
normalisation parameter. Consequently, we re-performed a fit of the COMPASS data 
by adding to the multiplicities of Eq.~(\ref{mult-gaus}) an overall multiplicative 
normalisation factor, linearly dependent on $y$, parameterised as 
\be
N_y = A + B\,y \,,
\label{y-dep}
\ee
which implies two additional parameters $A$ and $B$, assumed to be universal and 
flavour independent.  
 
With this parameterisation the quality of our best fit improves very significantly, 
resulting in a total $\chi^2_{\rm dof}$ of $3.42$ ($z<0.6$), corresponding to $A=1.06 
\pm 0.06$ and $B=-0.43 \pm 0.14$ and only very slightly different values of the 
Gaussian widths with respect to those previously obtained, Eq.~(\ref{compass-par}), 
as shown in the third entry of Table~\ref{tab:chi-sq-compass}.
A similar improvement is obtained for $z<0.7$.

\begin{figure}[t]
\includegraphics[width=0.8\textwidth, angle=-90]{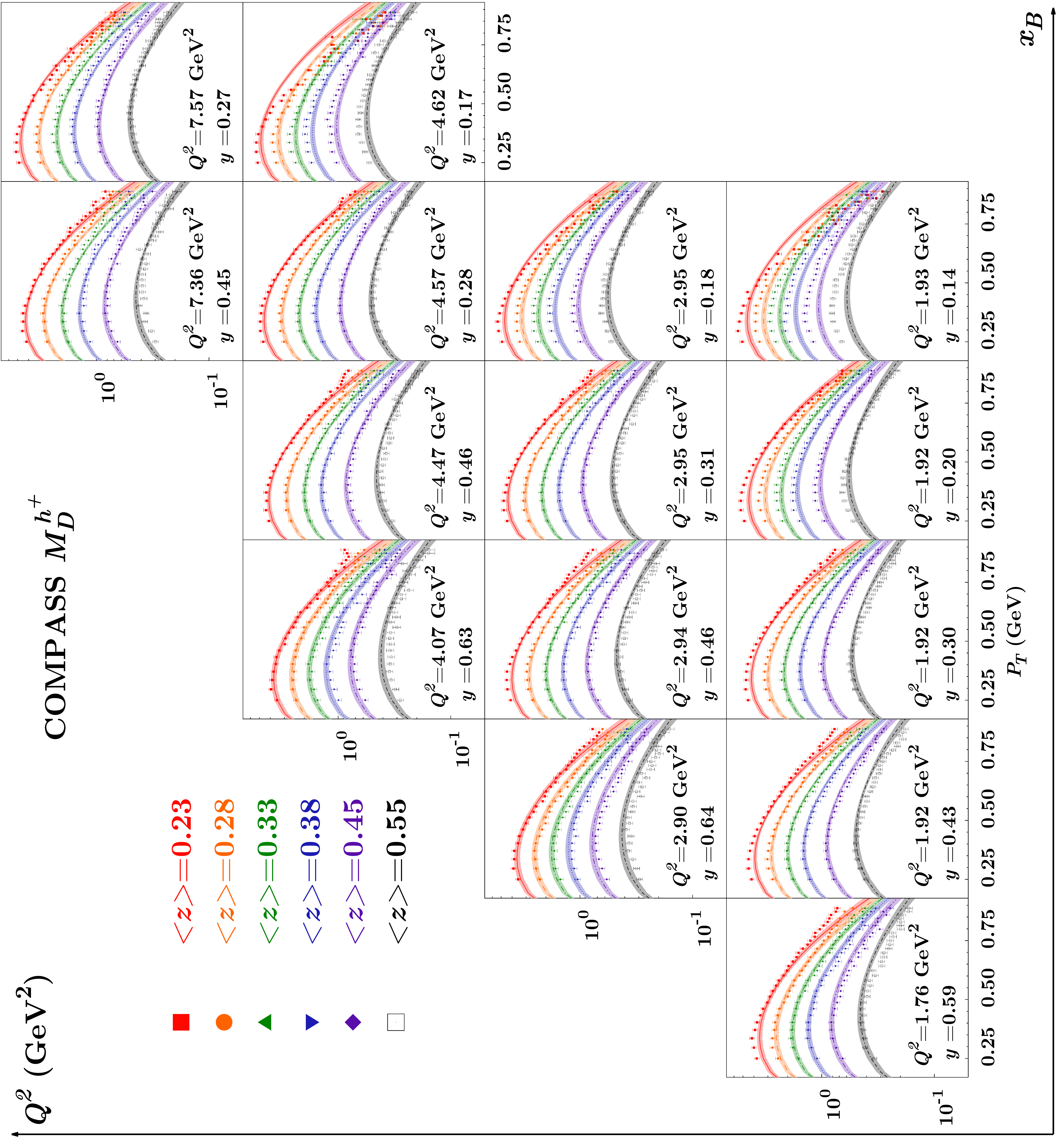}
\caption{\label{fig:compass-D-h+-y-dep}
The multiplicities obtained including the $y$-dependent normalisation factor of Eq.~(\ref{y-dep}) 
are compared with the COMPASS measurements for $h^+$ SIDIS 
production off a deuteron target. The shaded uncertainty bands correspond
to a $5$\% variation of the total $\chi^2$.}
\end{figure}

\begin{figure}[t]
\includegraphics[width=0.8\textwidth, angle=-90]{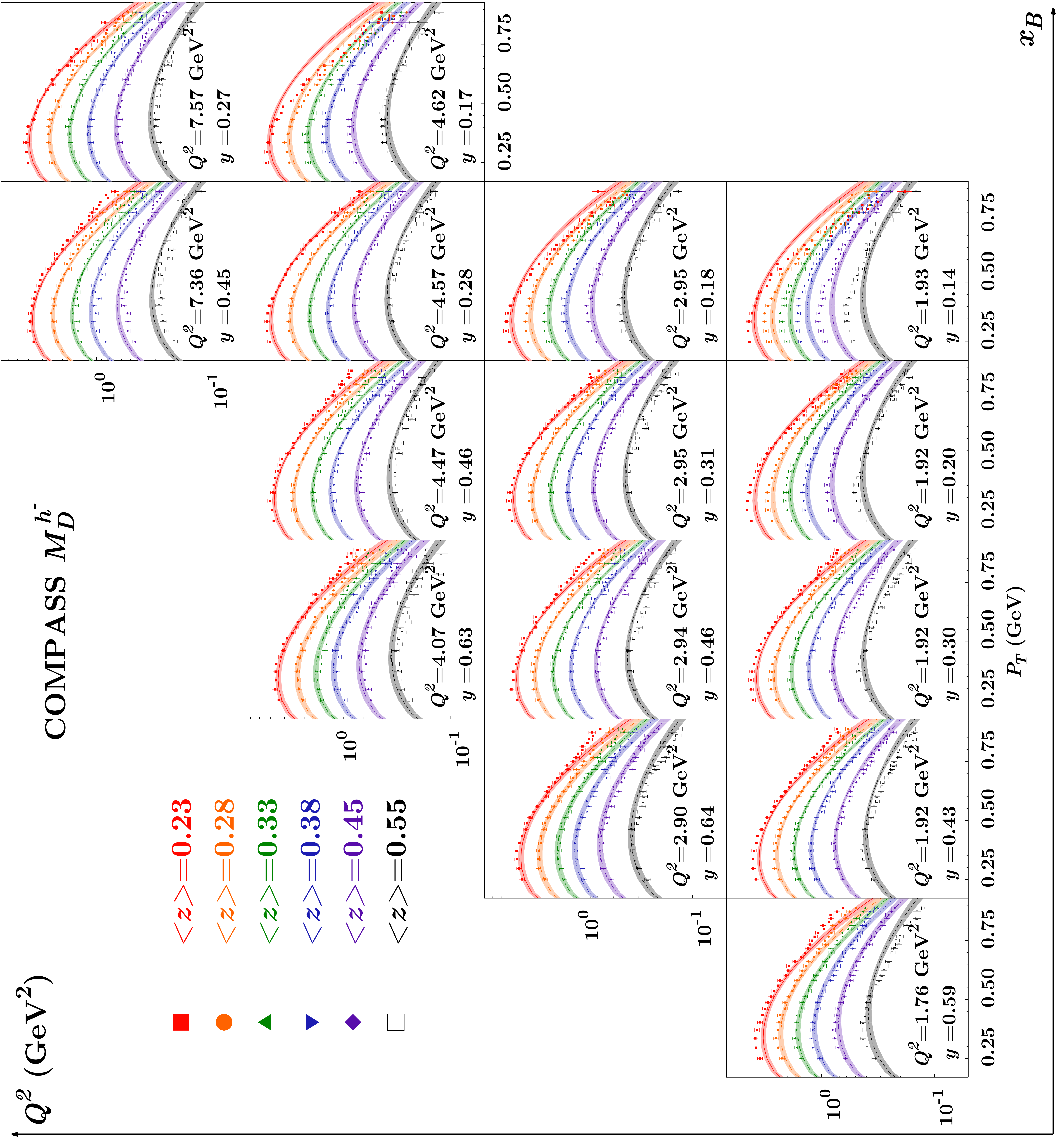}
\caption{\label{fig:compass-D-h--y-dep}
The multiplicities obtained including the $y$-dependent normalisation factor of Eq.~(\ref{y-dep}) are 
compared with the COMPASS measurements for $h^-$ SIDIS production off a deuteron target. 
The shaded uncertainty bands correspond
to a $5$\% variation of the total $\chi^2$.}
\end{figure}

The results of our best fit, for positively and negatively charged hadronic 
production, are presented in Fig.~\ref{fig:compass-D-h+-y-dep} 
and~\ref{fig:compass-D-h--y-dep} respectively. By comparing these figures with 
the analogous Figs.~\ref{fig:compass-D-h+} and \ref{fig:compass-D-h-}, in 
particular the plots on the left sides, corresponding to large $y$ (and low 
$\xb$) bins, one can see that the huge gain in $\chi^2$ is due to the fact that 
only with this second approach we can reproduce the normalisation of the data. 
Further comments on this issue will be made in the conclusions.

To complete our analysis, we finally performed a fit in which we allowed for 
two individual Gaussian widths for the favoured and disfavoured fragmentation 
functions, similarly to what was done for the HERMES measurements,
assuming that the final hadrons are mostly pions. 
However, in this case, these two parameters turn out to be roughly the same. 
For $z<0.6$, we find: $\avk = 0.60 \pm 0.15$, $\avp_{\rm fav} = 0.20 \pm 0.04$, 
$\avp_{\rm disf} = 0.20 \pm 0.05$, for the Gaussian widths and 
$A = 1.06 \pm 0.06$, $B = -0.43 \pm 0.14$, for the normalisation constants. 
These parameters correspond to $\chi^2_{\rm dof} = 3.42$. 

In conclusion, we have found that COMPASS data show the need for an overall 
$y$-dependent normalisation; having fixed that, then the multiplicities appear 
to be in agreement with a Gaussian dependence, although the resulting value 
of $\chi^2_{\rm dof}$ remains rather large.
Notice that this normalisation issue is not observed in the HERMES multiplicities 
and its origin, at present, cannot easily be explained
and deserves further studies. 

Some general comments on COMPASS results, inspired and guided by our grouping 
of the data in the panels of Figs.~\ref{fig:compass-D-h+} and 
\ref{fig:compass-D-h-} and by the study presented in Fig.~\ref{fig:N-y}, could 
help to understand the origin of the large values of $\chi^2_{\rm dof}$. 
Let us consider, for example, the data in the different panels of the same row 
in Fig.~\ref{fig:compass-D-h+}. The multiplicity data grouped there have all 
very similar values of $Q^2$ and are separated in bins of $z$; one can notice,
going from left to right, that data with very close value of $Q^2$ and $z$, still 
show a sharp $x$ dependence. This can hardly be reproduced by Eq.~(\ref{mult-gaus}),
even considering eventual higher order corrections. Similar considerations 
apply to Fig.~\ref{fig:compass-D-h-}.     

The large $\chi^2$ which persists even in the case in which we correct with
$N_y$, is mainly due to some particular subsets of data, as one can see from
Figs.~\ref{fig:compass-D-h+-y-dep} and \ref{fig:compass-D-h--y-dep} looking at
the rightmost lower panels. These data, if compared with those in the panels to
their immediate left (which have very similar values of the binned kinematical
variables) show a sudden sharp change, which our smooth Gaussian parameterisation
is unable to describe. Such a sharp change corresponds to the first, lowest
$y$ point, in Fig.~\ref{fig:N-y}.  

%
%

%
%

\section{Scale evolution}

In this Section we perform a series of simple tests to check whether the SIDIS 
multiplicity measurements show any indication of transverse momentum scale 
evolution. So far, in our model, the $Q^2$ scale evolution has been taken into 
account only through the DGLAP evolution of the collinear distribution and 
fragmentation functions, Eqs.~(\ref{unp-dist}) and~(\ref{unp-frag}), while no 
$Q^2$ dependence is inserted in the transverse momentum distributions. 

Multivariate measurements of SIDIS multiplicities offer a unique chance to study the 
scale evolution of their transverse momentum spectra. 
A complete description of this 
evolution should be done in the context of a full TMD evolution scheme, where 
much progress has recently been achieved~\cite{Ji:2004xq,Ji:2004wu,
Collins:2011zzd,Aybat:2011zv}. However, we do not use here the complete 
expression of the evolution equations, which are quite complex and can differ 
according to the scheme used; rather, we exploit some simple modelling which 
embeds only those features of the scale evolution which are most relevant in 
our studies. In particular we expect that, in the kinematical ranges covered 
by the HERMES and COMPASS experiments, the leading contribution should be given 
by the non-perturbative, model dependent, part of the TMD evolution equation, 
which consists essentially in a scale dependence of the PDF and FF Gaussian 
widths. 

For our SIDIS studies we choose a parameterization of $\avk$ and $\avp$, 
Eqs.~(\ref{unp-dist}) and (\ref{unp-frag}), inspired to that used for Drell-Yan 
and $e^+e^-$  processes~\cite{Nadolsky:2000ky,Landry:2002ix} in the context of 
the Collins-Soper-Sterman resummation scheme~\cite{Collins:1984kg}:
\bea
&&\avk =  g_1 + g_2 \ln(Q^2/Q^2_0) + g_3 \ln(10 \, e \, x)
\label{evol-gaussians1}\\
&&\avp = g^\prime _1 + z^2 g^\prime_2 \ln(Q^2/Q^2_0)\,,
\label{evol-gaussians2}
\eea
with $g_1$, $g_2$, $g_3$, $g^\prime_1$, and $g_2^\prime$ free parameters to be 
extracted by fitting the experimental data. Incidentally, the choice of the 
argument of $\ln(10\,e\,x)$ is just a simple way to assign to the $g_3$ 
parameter a suitable normalisation, such that $\avk = g_1 + g_3$ at $x=0.1$ 
and $Q^2=Q_0^2$. As already noticed, the SIDIS cross section is not sensitive 
to the individual contributions of $\avk$ and $\avp$, but only to their linear 
combination, $\avPT$, see Eqs.~(\ref{G-FUU}) and (\ref{avPT}). Therefore, we take
\be
\avPT = g^\prime _1 + z^2 [ g_1 + g_2 \ln(Q^2/Q^2_0) + g_3 \ln(10\,e\,x)]\,,
\label{evol-PT}
\ee
where the $g_2$ parameter used here replaces the sum of $g_2$ and  $g_2^\prime$ 
in Eq.~(\ref{evol-gaussians1}) and (\ref{evol-gaussians2}). Notice that this is 
the reason why the value of $g_2$ extracted by fitting SIDIS data, in general, 
cannot be directly applied to other processes, like Drell-Yan or $e^+e^-$ 
scattering, where knowledge on the individual contribution of either the 
distribution or the fragmentation TMD Gaussian width is needed. 

For HERMES data we found no sensitivity to these parameters, confirming the 
good agreement of their measured multiplicities with the most simple version of 
our Gaussian model.

The description of COMPASS data is much more involved. First of all, we have to 
introduce the $y$-dependent normalisation, as in Eq.~(\ref{y-dep}); then, we study 
the scale dependence of the $\avPT$ Gaussian width. We find that this width is not 
constant and can actually depend on $Q^2$ and $x$: by allowing the dependence of  
Eq.~(\ref{evol-PT}), the obtained $\chi^2_{\rm dof}$ decreases from $3.42$ to 
$2.69$. However, we realised that a better improvement can be achieved by  
choosing, for $\avPT$, the parametric form 
\be 
\avPT =  a^\prime _1 +  a^\prime _2 \ln(10 \, y) + z^2 [ a_1 + a_2 \ln(10 \, y)]\,,
\label{evol-PT-y}
\ee
which amounts to replacing the dependence on $Q^2$ and $x$ by a dependence on 
their ratio $y=Q^2/(x\,s)$. In this case we obtain $\chi^2_{\rm dof} = 2.00$. 
A further decrease in the total $\chi^2$ can be achieved by adding a square 
root term, $g_{3f} \,\sqrt{y}$, in the Gaussian width of the fragmentation 
function: in this case we obtain $\chi^2_{\rm dof} = 1.81$, close to the value 
found for the HERMES data, indicating a significant improvement with respect 
to the results obtained without scale dependence.  

Although such an ad-hoc $y$-dependence can only be considered as a phenomenological 
attempt to improve the fits, it is consistent with the $W^2$ dependence 
of $\avPT$ already found by the COMPASS Collaboration itself in their 
data, see Fig.~$9$ of Ref.~\cite{Adolph:2013stb}, and by the EMC Collaboration,
see Figs.~6 and 8 of Ref.~\cite{Ashman:1991cj}.

\section{Comparison with previous extractions and other experiments \label{error}}

The SIDIS multiplicity data used in our present fits result from the most 
recent analyses of the HERMES and COMPASS Collaborations. 
They represent, so far, the only multivariate analyses available. 

Additional measurements are provided by the early EMC results of 
Ref.~\cite{Ashman:1991cj} or by the more recent SIDIS studies of JLab 
CLAS~\cite{Osipenko:2008aa} and HALL-C~\cite{Mkrtchyan:2007sr,Asaturyan:2011mq} 
Collaborations. As we will explain below, these data are not best suited for the 
extraction of the free parameters of our fit and we have not used them. 
However, it is worth and interesting to check whether or not the parameters 
extracted here are consistent with the available EMC and JLab measurements. 

\begin{figure}[t]
\includegraphics[width=0.32\textwidth, angle=-90]{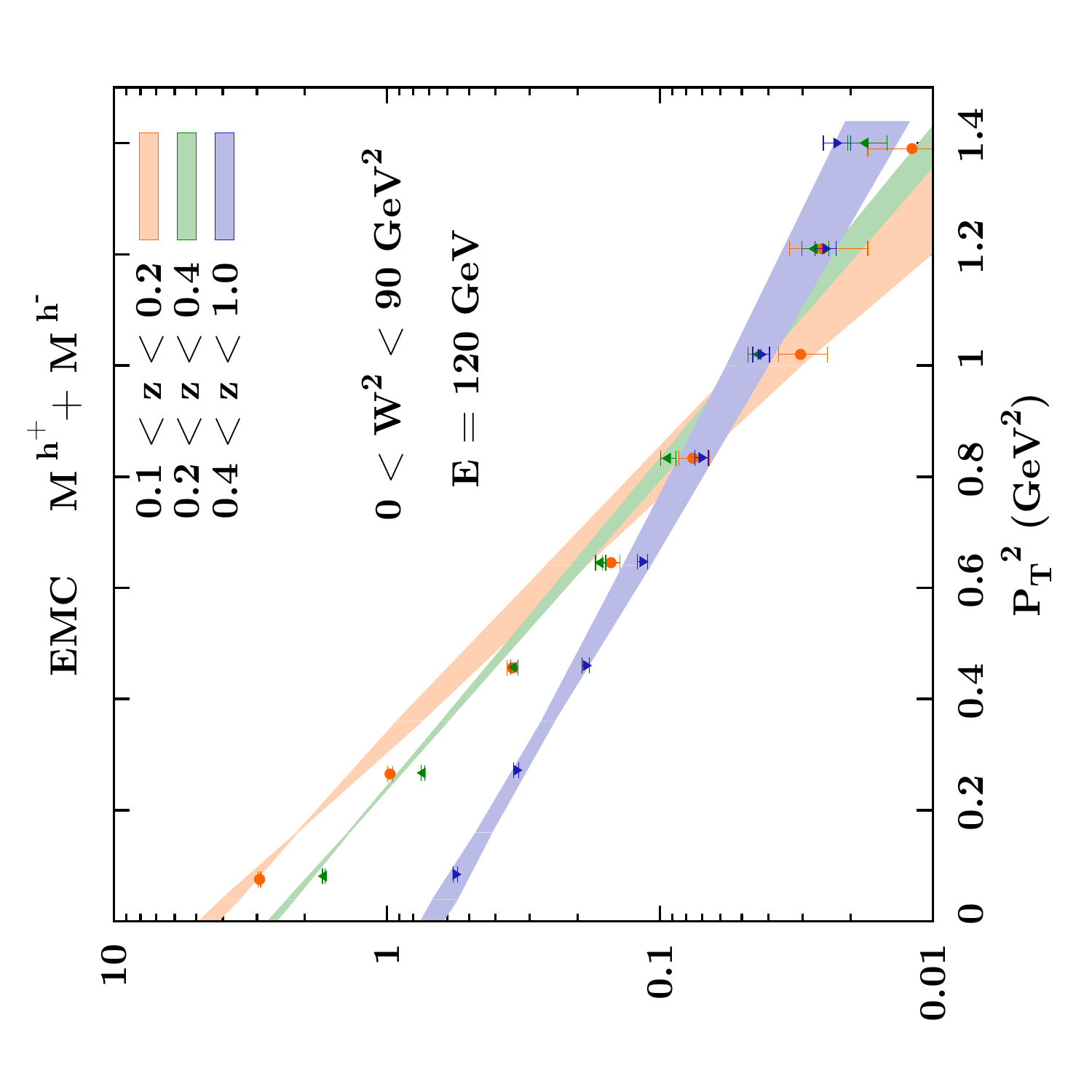}
\includegraphics[width=0.32\textwidth, angle=-90]{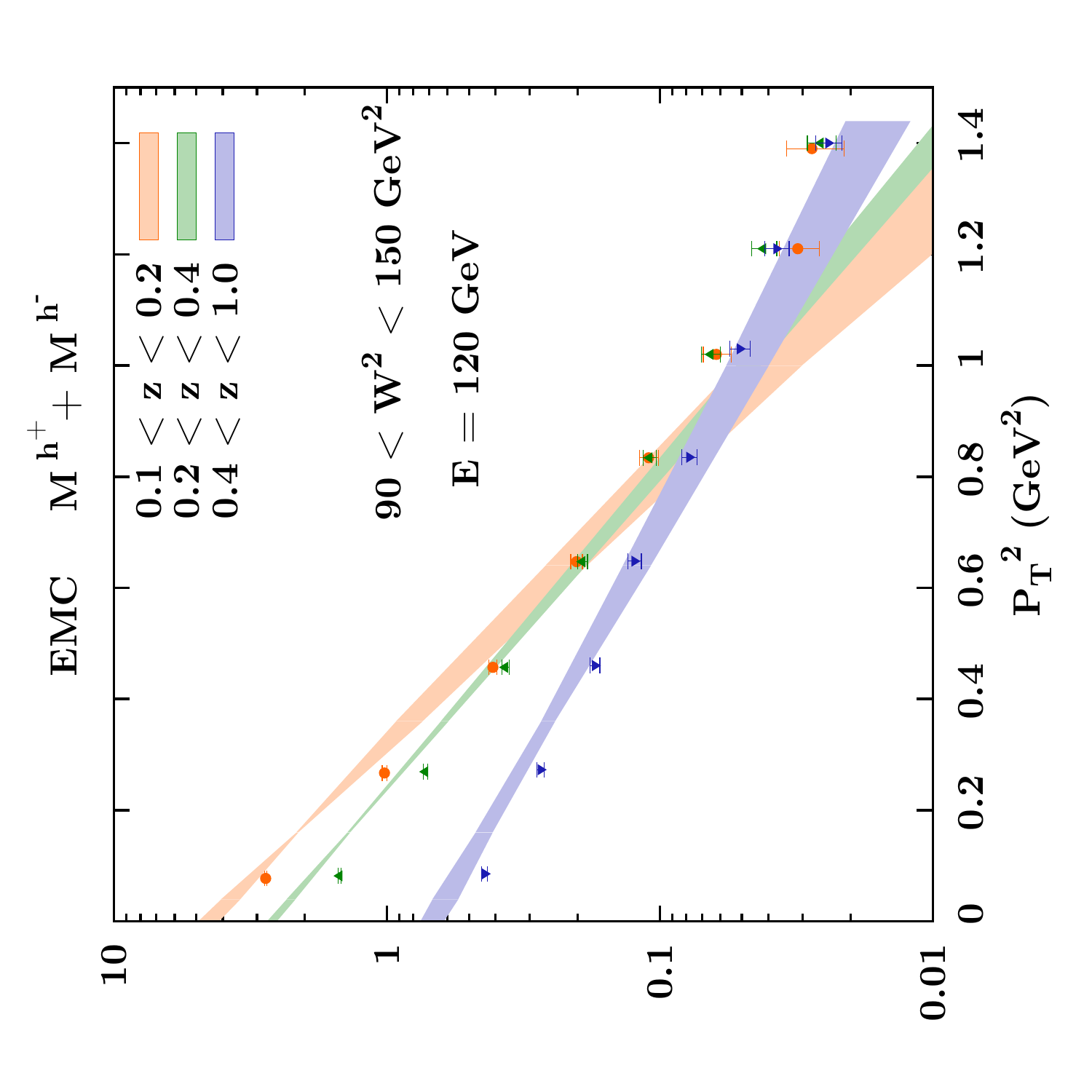}
\includegraphics[width=0.32\textwidth, angle=-90]{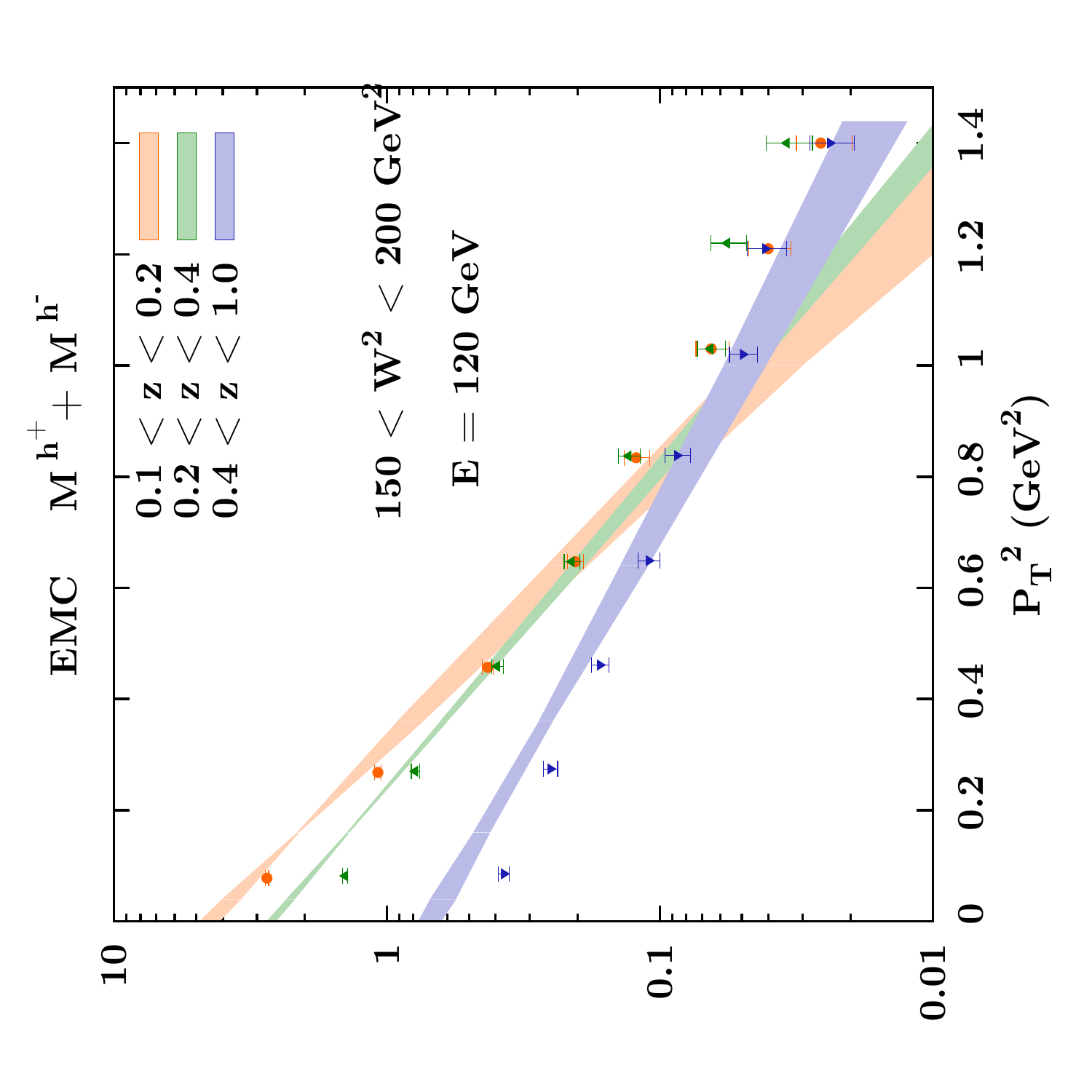}\\
\includegraphics[width=0.32\textwidth, angle=-90]{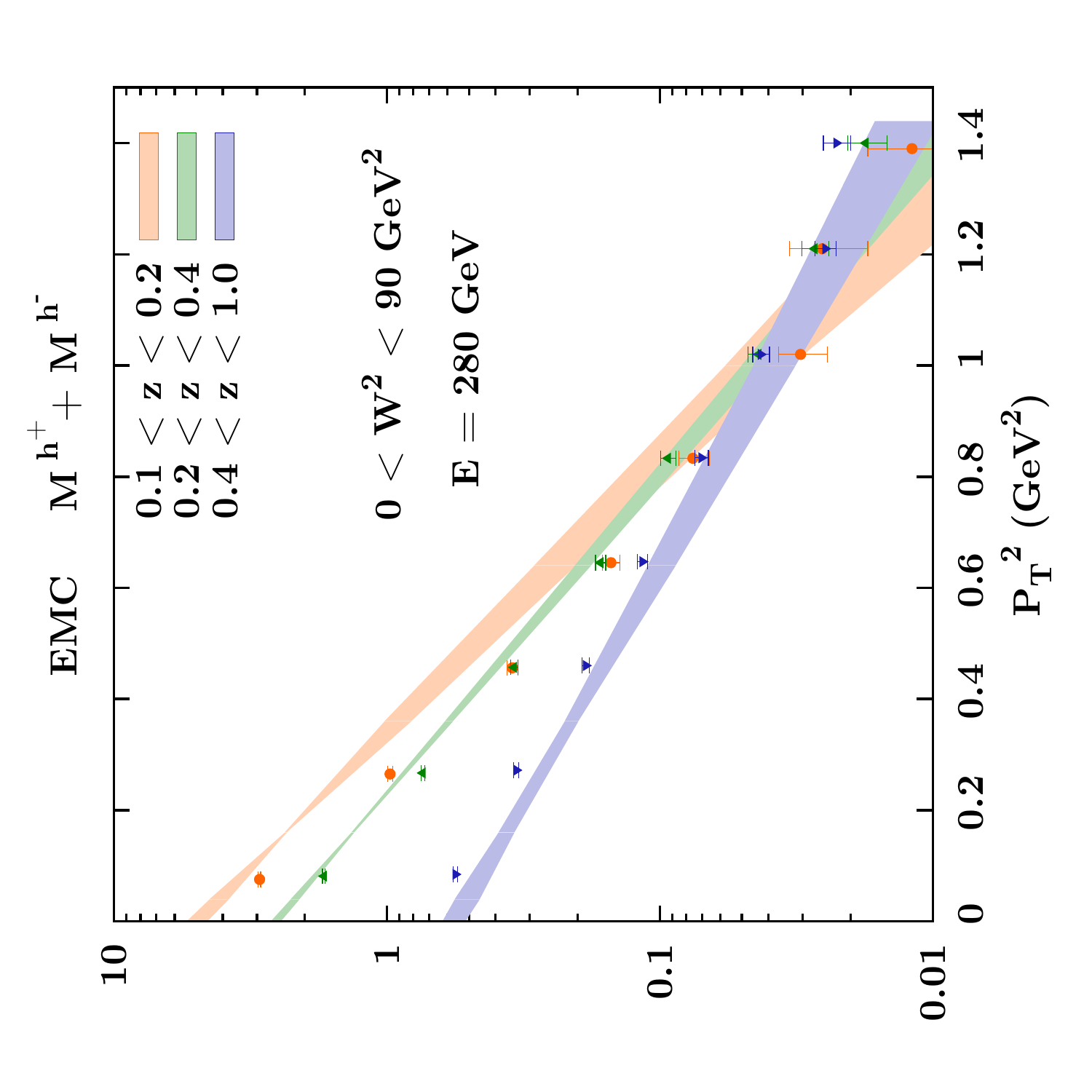}
\includegraphics[width=0.32\textwidth, angle=-90]{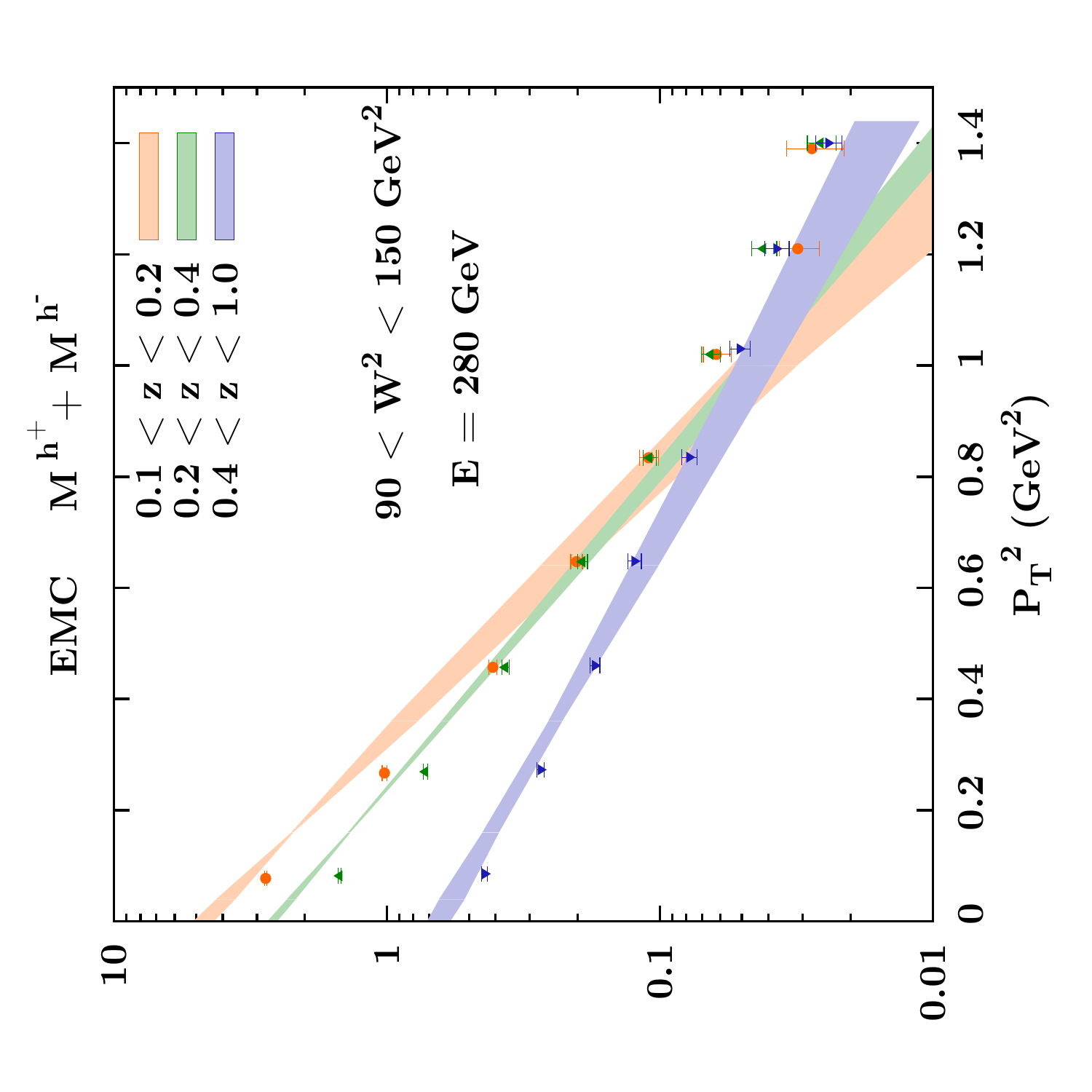}
\includegraphics[width=0.32\textwidth, angle=-90]{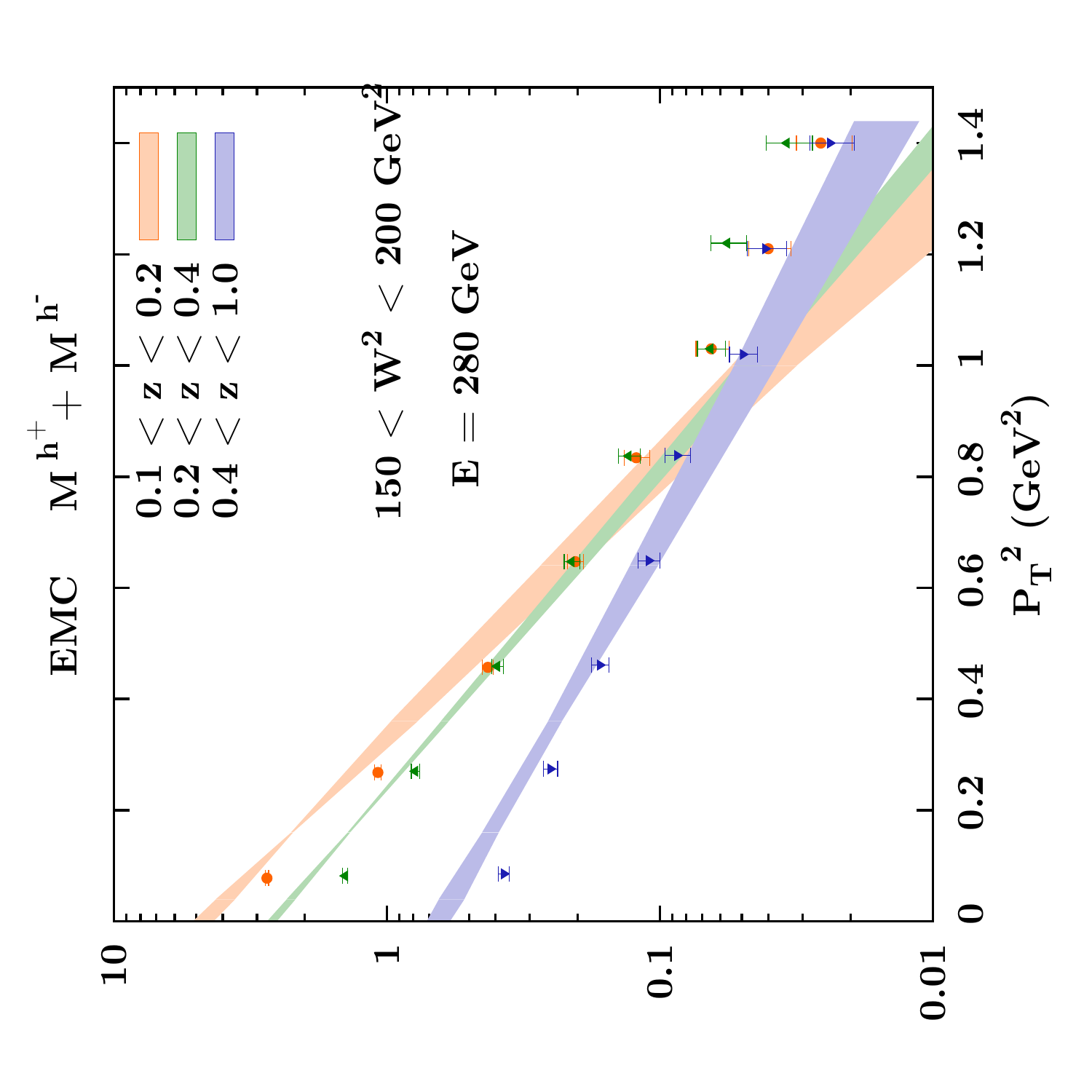}
\caption{\label{fig:EMC}
The EMC multiplicities~\cite{Ashman:1991cj}  are plotted as functions of $P_T^2$ for three bins of 
$z$, $0.1<z<0.2$, $0.2<z<0.4$ and $0.4<z<1.0$, 
and of invariant mass, $W^2<90$, $90<W^2<150$ and $150<W^2<200$. 
These data are compared to the predictions, computed at two different beam energies, 
$E_{lab}=120$ GeV (upper panel) and $E_{lab}=280$ (lower panel), 
by using the $\avk$ and $\avp$ extracted 
from the best fit of the COMPASS multiplicities (see the second entry of 
Table~\ref{tab:chi-sq-compass})      . 
}
\end{figure}

\begin{figure}[t]
\includegraphics[width=0.7\textwidth, angle=-90]{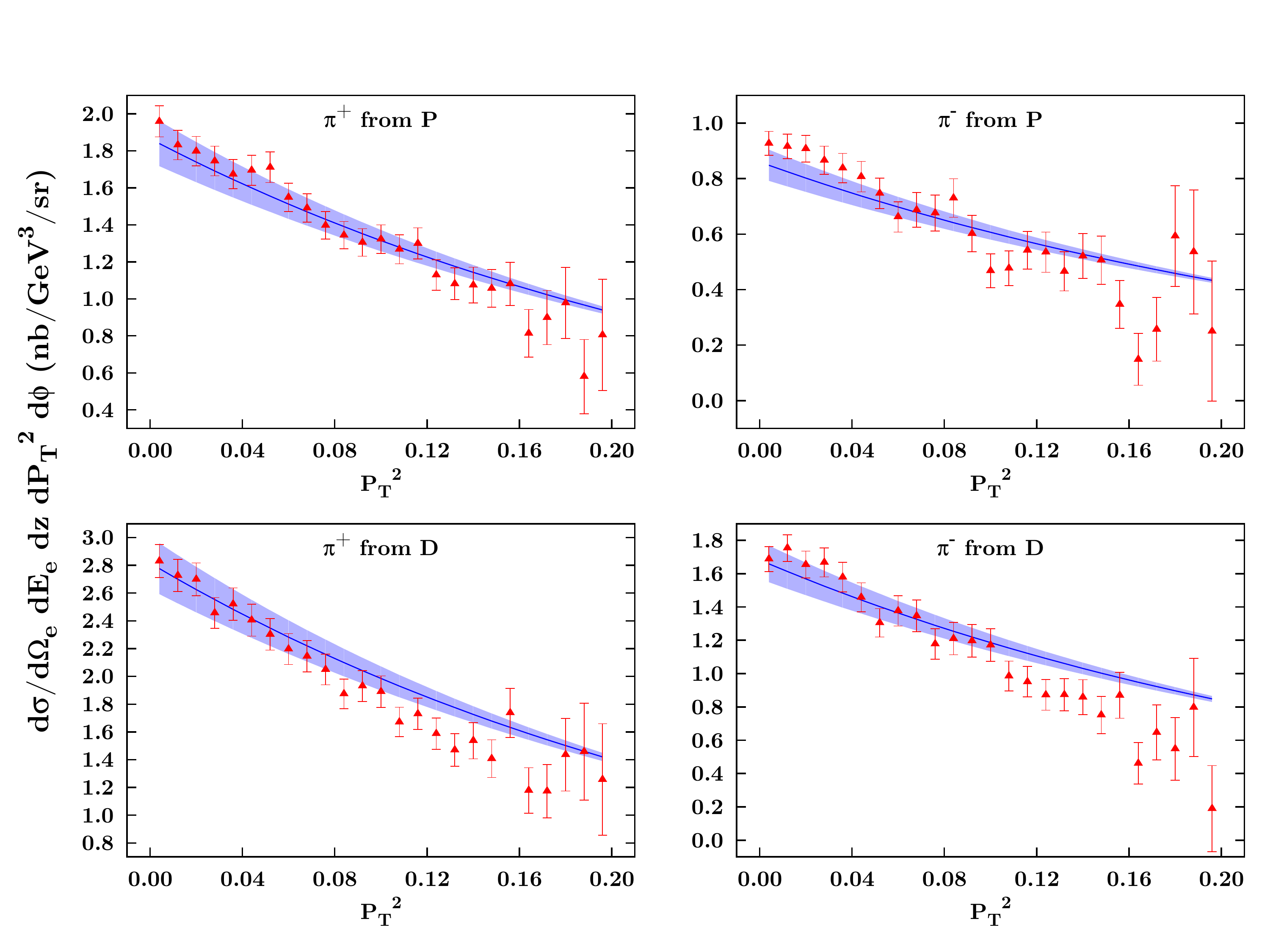}
\caption{\label{fig:jlab1}
JLab-Hall C $P_T$-distributions for $\pi^+$ and $\pi^-$ production from SIDIS scattering off a 
$^1H$ (proton) and $^2H$ (deuteron) targets, plotted as a function of $P_T^2$. 
Experimental data, corrected for vector meson production, are from Ref.~\cite{Asaturyan:2011mq}.
The lines are the prediction of our model obtained by using the parameters extracted from a best 
fit of the HERMES multiplicities, see Table~\ref{tab:chi-sq-hermes}.
The shaded uncertainty bands are computed by propagating the error on the two free 
parameters, as explained in Appendix~\ref{error}.}
\end{figure}

\begin{figure}[t]
\includegraphics[width=0.4\textwidth, angle=-90]{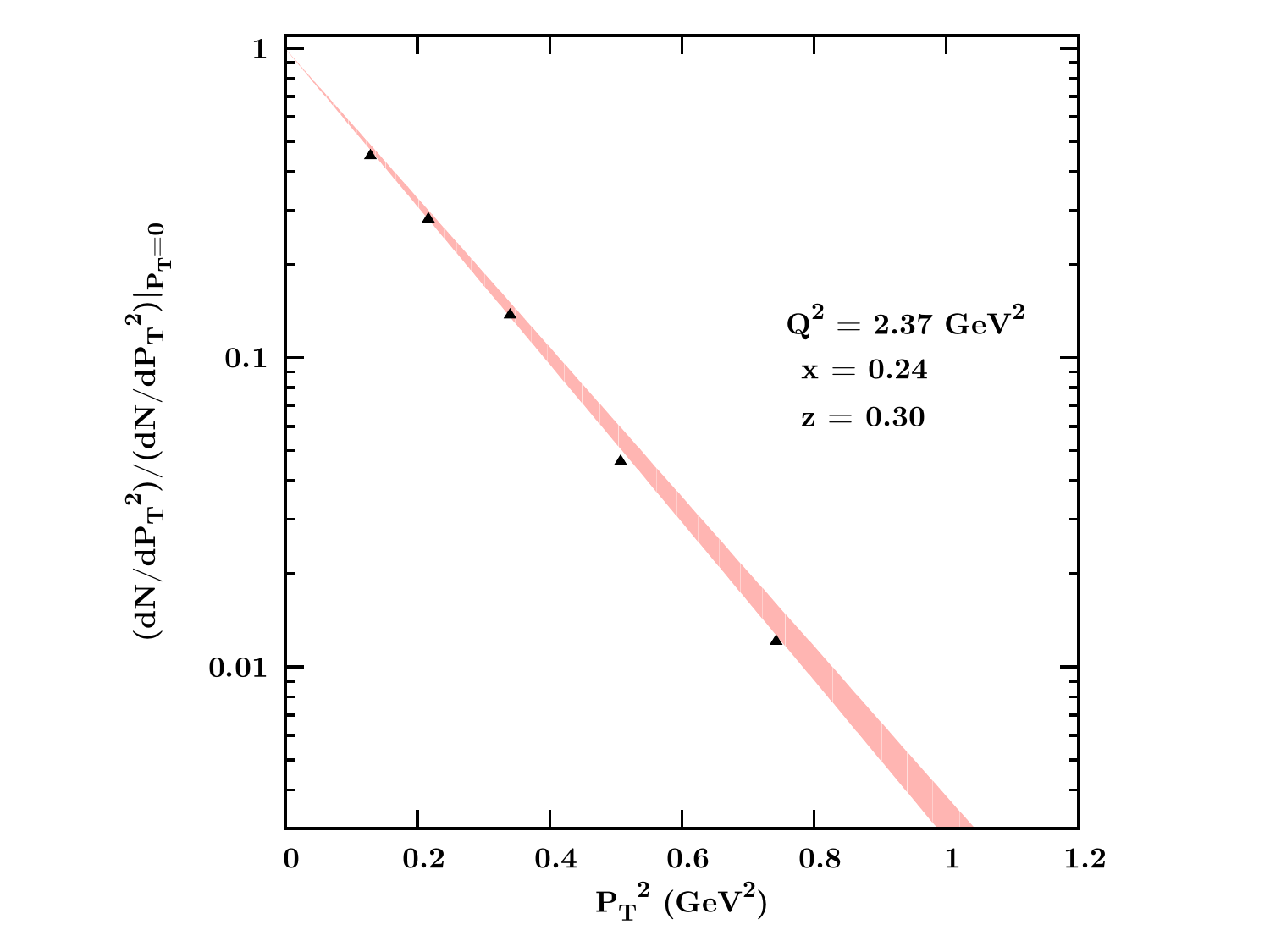}
\caption{\label{fig:jlab4}
JLab-CLAS~\cite{Osipenko:2008aa} 
$(\epsilon {\cal H}_1 + {\cal H}_2)(P_T)/(\epsilon {\cal H}_1 + {\cal H}_2)(P_T=0)$ 
for $\pi^+$ and $\pi^-$ production from SIDIS scattering off 
proton and deuteron targets, are compared to the predictions obtained from our model 
by using the parameters extracted from the best fit of the HERMES multiplicities, 
see Table~\ref{tab:chi-sq-hermes}. 
}
\end{figure}

\begin{itemize}
 \item 
The EMC Collaboration~\cite{Ashman:1991cj} measured
$P_T$-distributions in eleven different runs presented in one merged data set, 
averaging over four different beam energies, three different nuclear targets, 
without any identification of the final hadrons (not even their charges), 
and arranging the data in three different bins of $z$ and several ranges of $W^2$.  
In Ref.~\cite{Anselmino:2005nn} we exploited these measurements, 
together with the EMC measurements~\cite{Arneodo:1986cf} of the azimuthal 
dependence of the SIDIS cross section, for a preliminary study of the Gaussian 
widths of the unpolarised distribution and fragmentation functions. The values 
found there are slightly different from those we determine in the present fit.
Fig.~\ref{fig:EMC} shows the EMC multiplicities~\cite{Ashman:1991cj} as functions 
of $P_T^2$, for three bins of $z$, $0.1<z<0.2$, $0.2<z<0.4$ and $0.4<z<1.0$, 
and of the invariant mass, $W^2<90$, $90<W^2<150$ and $150<W^2<200$ (in GeV$^2$). 
These data are compared with our predictions, computed at two different beam 
energies, $E_{lab}=120$ GeV and $E_{lab}=280$, using the values of $\avk$ and 
$\avp$ extracted from the best fit of the COMPASS multiplicities (see the second 
entry of Table~\ref{tab:chi-sq-compass}). The slope of our $P_T^2$ distributions, 
which represents the Gaussian width $\avPT$, qualitatively agrees with the EMC 
measurements. The merging of these measurements in {\it one} data set, makes 
any quantitative conclusion quite lax (as one can see by comparing, for instance, 
the upper and lower panels of Fig.~\ref{fig:EMC} which refer to two very different energies).
\item
Jlab-HALL-C Collaboration~\cite{Asaturyan:2011mq} provides unpolarised cross 
sections, for SIDIS scattering off $^1H$ (proton) and $^2H$ (deuteron) targets 
and for $\pi^+$ and $\pi^-$ final production, as a function of $P_T^2$, at 
fixed values of $x=0.32$, $Q^2=3.2$ GeV$^2$ and $z=0.55$. 
At fixed $z$, these distributions are only sensitive to 
the total Gaussian width $\avPT=\avp +z^2 \avk$, and not to the separate 
values of $\avkt$ and $\avp$. In Fig.~\ref{fig:jlab1} we compare these 
JLab measurements~\cite{Asaturyan:2011mq} with the 
$P_T$ distribution calculated in our scheme by using the parameters 
extracted from a best fit of the HERMES multiplicities, Eq.~(\ref{hermes-par}). 
We use these parameter values as the kinematical region explored at JLab is  
similar to that spanned in the HERMES experiment. 
An overall factor $N=1.5$, common to all four data sets, has been introduced.
Notice that $\avPT$, which here represents the only parameter to which these 
data are sensitive, appears both as the distribution width and as a normalisation factor, see Eq.~(\ref{mult-gaus}). 
At a qualitative level, we can conclude that the Gaussian model and the extracted 
values of $\avkt$ and $\avp$ are not in conflict with the JLab cross section 
measurements of Ref.~\cite{Asaturyan:2011mq}. However, different normalizations 
could require different widths.
The shaded uncertainty bands are computed by propagating the error on the 
two free parameters, as explained in Appendix~\ref{bands}. Notice that the data 
extend to a region ($P_T^2 < 0.04$ GeV$^2$) excluded in our best fit analysis. 

\item
In Fig.~\ref{fig:jlab4} we show the CLAS extraction of  
$(\epsilon {\cal H}_1 + {\cal H}_2)(P_T)/(\epsilon {\cal H}_1 + {\cal H}_2)(P_T=0)$, 
defined in Eqs.~(1) of Ref.~\cite{Osipenko:2008aa}, as a function of $P_T^2$ 
at the fixed values of $x=0.24$ and $z=0.30$; in our model, Eq.~(\ref{mult-gaus}), 
this ratio corresponds to the actual Gaussian $e^{-P_T^2/\avPT}$. From 
the theoretical point of view, this quantity is more suitable for a 
comparison with our predictions, as it is unambiguously normalised. 
However, from the experimental point of view, its determination requires the extrapolation of $(\epsilon {\cal H}_1 + {\cal H}_2)(P_T)$ down to $P_T=0$.
The plot shows that the line we obtain by using the 
Gaussian widths extracted from HERMES multiplicities is qualitatively in 
agreement with the data taken at $Q^2=2.37$, showing the right slope. 
Furthermore, CLAS finds quite a strong scale dependence (not shown here), as 
their slopes change considerably at $Q^2=2.00$ and $Q^2=1.74$, 
a dependence which we cannot account for in our scheme. Instead, differences 
among slopes at different values of $Q^2$ could be obtained by applying the 
appropriate kinematical cuts to the integration over $\kt$, rather than 
integrating over the full $\kt$ range, as suggested, for example, in 
Ref.~\cite{Boglione:2011wm}. 

\end{itemize}
%
%
%
%
%
%
\section{Conclusions}
We have considered a large amount of unpolarised SIDIS multiplicity data,
recently made available by the HERMES~\cite{Airapetian:2012ki} and
COMPASS~\cite{Adolph:2013stb} Collaborations, and their $P_T$ dependence.
Such a dependence, in selected kinematical regions, is believed to be
generated by the intrinsic motion of quarks in a nucleon (information encoded
in the TMD-PDFs), and of the hadron in the fragmentation process (encoded in
the TMD-FFs). The separation of the data in different bins of $x$, $Q^2$,
$z$ and $P_T$ allows to study the $P_T$ dependence in appropriate
independent kinematical regions, obtaining safe information on the TMDs.

We have parameterised the unknown TMDs in a most simple way,
Eqs.~(\ref{unp-dist}) and (\ref{unp-frag}).\\
Our results can be summarised as follows.

\begin{itemize}
\item
{\it HERMES multiplicities}.

Our simple Gaussian parameterisation delivers a satisfactory description of the
HERMES data points over large ranges of $x$, $z$, $P_T$ and $Q^2$, selected
according to Eqs.~(\ref{kin-cuts1}) and (\ref{kin-cuts2}). These measurements
are well described by a TMD Gaussian model with constant and flavour independent
widths, $\avk$ and $\avp$, which we extract as (the only two) free parameters
of our fit. There is no need of an overall normalisation. HERMES multiplicities 
do not show any significant sensitivity to additional
free parameters: the fits do not improve by introducing a $z$-dependence in the
Gaussian widths of the TMD-FFs or by allowing a flavour dependence in the
Gaussian widths of the TMD-PDFs. We only find a slight improvement in $\chi^2$
by using different (constant) Gaussian widths in the TMD-FFs; the disfavoured
fragmentation functions show a preference for a width slightly
wider than that of the favoured fragmentation functions.

\item
{\it COMPASS multiplicities}.

Fitting COMPASS data turns out to be more difficult. One would expect their
high statistics and fine accuracy to provide strong constraints on our model.
However, while the Gaussian shape of the $P_T$ dependence is qualitatively
well reproduced, there are some unresolved issues with their relative overall
normalisation. By performing a bin-by-bin analysis, we realised that
different normalisation constants are required for different $y$-bins.
In particular, we found that this $N_y$ normalisation factor is roughly
$1$ for very small $y$, while decreasing linearly with growing $y$:
this implies almost a factor two difference between the largest and the
smallest $y$ bin, even within very close values of $Q^2$ and $x$, which would be
very difficult to accommodate in a QCD driven scheme, even considering scale evolution.
This can clearly be seen, for instance, by comparing the panels of each row in
Fig.~\ref{fig:compass-D-h+} going from left to right.
We are presently unable to determine the origin of this effect, which indeed needs
further investigation, both on the theoretical and experimental sides.

The COMPASS fit returns a $\avp$ TMD-FF Gaussian width slightly larger than
that extracted from the HERMES multiplicities, while it delivers similar $\avk$
values. Comments on the scale evolution sensitivity of COMPASS multiplicitites
will be presented in the next item.

Notice that this analysis has been performed on the
$2004$ run data, when the COMPASS detector was not yet completely set up and no
RICH was installed for final hadron separation. Future analyses of more recent
COMPASS data with hadron identification might help to clarify the situation.

\item
{\it Scale evolution}.

In our fit, with the phenomenological
parameterisation of Eqs.~(\ref{unp-dist}) and~(\ref{unp-frag}), the only
dependence on $Q^2$ is included in the collinear part of the TMD, {\it i.e.}
in the collinear PDF or FF factor. The width of the Gaussian, which gives the
$\kt$ ($\pp$) dependence of the TMDs, does not include any scale dependence.
Moreover, the Gaussians are normalised to one (if one integrates over the full
$\kt$ range) and consequently our $P_T$ integrated cross section corresponds to
the usual LO collinear SIDIS cross section.

As an alternative, while retaining this normalisation, we have tried new
parameterisations which allow for a $Q^2$ and/or $x$-dependence of the Gaussian
widths, Eq.~(\ref{evol-PT}), or even some $y$-dependence, Eq.~(\ref{evol-PT-y}).
For the HERMES data we did not find any significative $x$ or $Q^2$ dependence
in the trasnverse momentum spectra.\\
For the COMPASS data, instead, some improvement in the quality of the fit can
actually be obtained. However, due to the unresolved issues discussed above,
we cannot give a clear interpretation of this sensitivity
and prefer not to draw, at this stage, any definite conclusion.

Several general considerations should not be forgotten. It is quite possible that
the span in $Q^2$ of the available SIDIS data is not yet large enough to perform
a safe analysis of TMD evolution based only on these data. Another related issue
is that, always considering the SIDIS data set, the values of $P_T$, while being
safely low, are sometimes close to $Q$ and corrections to the TMD factorisation
scheme might be still relevant.

Only a combined analysis of SIDIS and Drell-Yan data, that would span in $Q^2$
from $1$ to approximately $80$ GeV$^2$, would allow a comprehensive study. Such
an analysis is beyond the scope of this paper and will be published elsewhere.

\item
{\it Comparison with other measurements}.

We have compared the multiplicities and $P_T$-distributions obtained by using
the parameters extracted from our best fit with EMC~\cite{Ashman:1991cj} and
JLab~\cite{Osipenko:2008aa,Asaturyan:2011mq} experimental data.
These data are not best suited for fitting and only a qualitative comparison
is possible. We found that EMC data are described reasonably well using the
parameter values extracted from COMPASS multiplicities while the JLab data
seem to be compatible with predictions based on the parameters extracted
from HERMES.
\end{itemize}

Summarising, from this study we find that the Gaussian widths describing the
$\kt$ distribution of the unpolarised TMD PDFs and the $\pp$ distribution of 
the unpolarize TMD FFs span respectively the approximate ranges
\be 
0.4 \lsim \avk \lsim 0.8 \>{\rm GeV}^2
\quad\quad\quad 
0.1 \lsim \avp \lsim 0.2 \>{\rm GeV}^2 \>.
\ee
Indeed, the actual values found here, as in Ref.~\cite{Signori:2013mda}, based 
on unpolarised multiplicity data, still have large intrinsic uncertainties.

Once again, it is important to point out that multiplicities are sensitive
to $\avPT$ only, and consequently they do not strictly allow a separate
determination of $\avk$ and $\avp$. Although the relation
\mbox{$\bfP _T =z\,\bfk _\perp + \bfp _\perp$}
is dictated by kinematics and is therefore model indipendent,
our relation between $\avp$, $\avk$ and $\avPT$, Eq.~(\ref{avPT}),
does depend on the parametrization chosen.

While the Gaussian dependence of the TMDs is well in agreement with multiplicity 
data, a precise determination of the separate values of $\avk$ and $\avp$ would 
require the analysis of other data, like the azimuthal dependences: in the Cahn 
effect, for instance, we are sensitive to the ratio 
$\avk/\avPT$~\cite{Anselmino:2005nn}. The study of azimuthal dependences, 
for the new HERMES and COMPASS data, will be performed in a forthcoming paper.

\section{Acknowledgements}

We thank V. Barone, U. D'Alesio, F. Murgia for useful discussions and A. Bressan,
A. Martin, M. Osipenko for providing their data tables and helping us to interpret
their measurements correctly. \\
A. P. work is supported by U.S. Department of Energy
Contract No. DE-AC05-06OR23177. \\
M.A. and M.B. acknowledge support from the European Community under the FP7
``Capacities - Research Infrastructures" program (HadronPhysics3, Grant Agreement 
283286) \\
M.A, M.B. and S.M. acknowledge support from the ``Progetto di Ricerca Ateneo/CSP" (codice TO-Call3-2012-0103). 

\appendix
\section{\label{bands}Uncertainty bands}

The uncertainty bands shown in the figures throughout this paper are computed
in terms of the covariance error matrix $C$ following the standard procedure,
shown for example in Ref.~\cite{Barone:2006xj}. In general, if we interpret
the free parameters of our fit, $p_i$, as the components of the vector $\bfp$,
the width of the uncertainty band corresponding to a generic function of some
kinematical variables $\bfx$, $f(\bfx;\bfp)$, will be given in terms of the
variation $\bfDelta_p (\bfx)$ of the value of each parameter with respect to
its value at the minimum $\chi^2$, $\bfp_0$:
\be
\Delta f (\bfx;\bfp _0)=|f (\bfx;\bfp_0 + \bfDelta _p (\bfx)) -
                               f (\bfx;\bfp_0 - \bfDelta_p (\bfx))|\,,
\ee
where the vector $\bfDelta_p (\bfx)\equiv (\Delta_{p_1}(\bfx), ...,
\Delta_{p_n}(\bfx)) $ is given by
\be
\bfDelta_p (\bfx) = \frac{C\,\bfpartial_p f(\bfx;\bfp)}
                        {\sqrt{\bfpartial _p f(\bfx;\bfp)\, C\, \bfpartial_p^T f(\bfx;\bfp)}}\,,
\ee
with
\be
C^{-1}_{ij} = \frac{1}{2} \frac{\partial ^2 \chi^2 }{\partial p_i \partial p_j}
\qquad {\rm and} \qquad
\bfpartial_p \equiv (\partial / \partial_{p_1}, ...,\partial / \partial_{p_n})\,.
\ee


\end{document}